\UseRawInputEncoding

\documentclass[preprint,superscriptaddress]{revtex4-1}

\usepackage{graphicx}
\usepackage{latexsym}
\usepackage{amsmath}
\usepackage{amssymb}
\usepackage{amsfonts}
\usepackage{color}
\usepackage{bm}
\usepackage{verbatim}
\usepackage{pagecolor,lipsum}
\usepackage{float}
\usepackage{hyperref}
\usepackage[normalem]{ulem}
\hypersetup{colorlinks=true,urlcolor=blue,linkcolor=blue,citecolor=blue}
\usepackage{comment}
\usepackage{siunitx}
\usepackage{xcolor}

\makeatletter
\renewcommand{\p@subsection}{}
\renewcommand{\p@subsubsection}{}
\makeatother

\begin{document}

\title{Time-reversal symmetry breaking and inherent Josephson diodes in a multi-component order parameter superconductor}

\date{\today}

\author{Remko Fermin}
\affiliation{Department of Materials Science \& Metallurgy, University of Cambridge, 27 Charles Babbage Road, Cambridge CB3 0FS, United Kingdom}
\affiliation{MESA+ Institute for Nanotechnology, University of Twente, 7500 AE Enschede, The Netherlands}
\author{Matthijs Rog}
\author{Guido Stam}
\affiliation{Huygens-Kamerlingh Onnes Laboratory, Leiden University, P.O. Box 9504, 2300 RA Leiden, The Netherlands}
\author{Daan Wielens}
\author{Joost Ridderbos}
\author{Chuan Li}
\affiliation{MESA+ Institute for Nanotechnology, University of Twente, 7500 AE Enschede, The Netherlands}
\author{Yoshiteru Maeno}
\affiliation{Toyota Riken – Kyoto University Research Center (TRiKUC), Kyoto 606-8501, Japan}
\author{Jan Aarts}
\author{Kaveh Lahabi}
\email{lahabi@physics.leidenuniv.nl}
\affiliation{Huygens-Kamerlingh Onnes Laboratory, Leiden University, P.O. Box 9504, 2300 RA Leiden, The Netherlands}

\begin{abstract}
Spontaneous time-reversal symmetry breaking is a defining feature of unconventional superconductivity and a prerequisite of the technologically relevant superconducting diode effect. However, its realization remains rare and exclusive to proximity-engineered weak links. Here we examine an alternative pathway to spontaneous time-reversal symmetry breaking by examining junctions that appear in single-crystal microstructures of the unconventional superconductor Sr$_2$RuO$_4$, lacking external superconducting proximity. We interpret the Josephson coupling as superconducting domains of a multi-component order parameter based on a plethora of anomalous effects --- including a field-free Josephson diode effect, fractional Shapiro steps, magnetic interference patterns deviating from the conventional flux quantum periodicity, and bias current-programmable bistable states. These microscopic devices corroborate earlier bulk probes of time-reversal symmetry breaking, giving a renewed impulse to the long-debated pairing symmetry of Sr$_2$RuO$_4$. Our results highlight superconducting domain walls as a platform for functional Josephson devices and the study of fundamental symmetry breaking.
\end{abstract}


 \maketitle

\section*{Introduction}

Spontaneous time-reversal symmetry breaking (TRSB) in superconductors is a hallmark of exotic macroscopic quantum states with additional degrees of freedom, such as spin or orbital angular momentum~\cite{wysokinski2019time,Ghosh_2021}. Several mechanisms can give rise to TRSB: chiral states with uncompensated orbital angular momentum~\cite{kallin2016chiral,wan_unconventional_2024}, spin-polarized triplet states with unequal spin populations~\cite{aoki_coexistence_2001,Ran_triplet,odd_freq_triplets,Linder2015}, and states with loop supercurrents~\cite{Ghosh_2021_loop,fittipaldi_unveiling_2021}.

TRSB has been mostly explored in engineered Josephson devices, where superconducting correlations are injected into carefully designed media with spin-orbit coupling and broken inversion symmetry. In these proximity systems, TRSB signatures manifest as the Josephson diode effect~\cite{ando_observation_2020,ubiquitous_diode,wu_field_free_2022,baumgartner_supercurrent_2022}, $\varphi$-junctions with double-well Josephson potentials~\cite{phi_junction_experiment}, $\varphi_0$-junctions with finite phase ground-states~\cite{szombati_josephson_2016,Konschelle_phi_0,strambini_josephson_2020}, or fractional Shapiro steps~\cite{Half_integer_ferro_1,trimble_josephson_2021}. The most unambiguous hallmark, however, is an asymmetry in the critical current interference pattern, $I_\text{c}(H)$, under reversal of both current and field direction --- a violation of $I_\text{c+}(H) = |I_\text{c-}(-H)|$. To date, such violation has only been realized in a handful of proximity junctions, often requiring electrostatic gating and external magnetic fields~\cite{CIT_Cuozzo,trimble_josephson_2021,TRS_not_broken_helical, wu_field_free_2022, szombati_josephson_2016, baumgartner_supercurrent_2022}.

 \begin{figure*}[h!]
 \centerline{$
 \begin{array}{c}
 \includegraphics[width=0.9\linewidth]{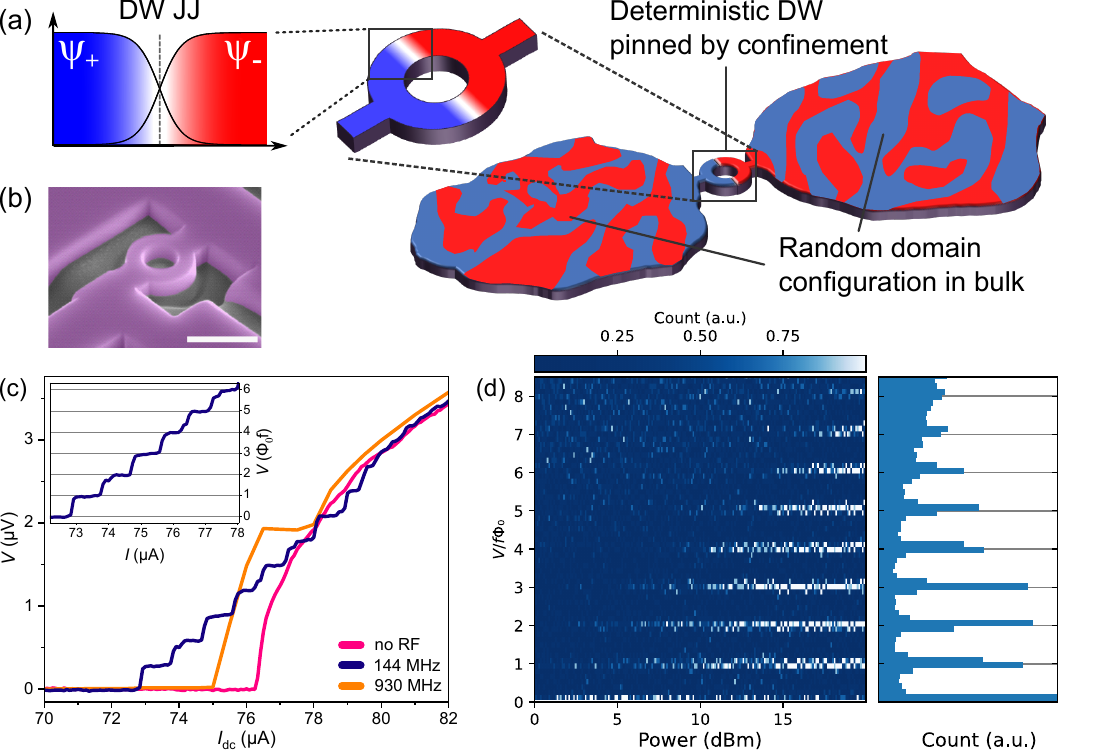}
 \end{array}$}
 \caption{\textbf{Josephson junction behavior in pristine Sr$_2$RuO$_4$ microstructures.} (a) Schematic of the proposed domain wall Josephson junctions (DW JJs) stable between two superconducting condensates (colored red and blue). The domain structure is randomly determined in the bulk, yet deterministically determined by the ring geometry~\cite{Yasui2020}. (b) A false-colored scanning electron micrograph of Ring 1. The crystal is colored purple; the device is capped by an additional layer of SiO$_\text{x}$ for protection against damages from the focused ion beam processing, which causes the dark contrast on top of the crystal; the scale bar corresponds to 1~{\textmu}m. (c) Current-voltage characteristics near the critical current ($I_\text{c}$) for different applied microwave frequencies, obtained on Ring 1 at 50 mK at maximum applied power (20 dBm). Clear quantized steps appear near $I_\text{c}$, the height of which linearly scales with the frequency of the applied microwave. The inset shows the data obtained at 144 MHz, rescaled to highlight that the step height matches the expected values for Shapiro steps. (d) The power dependence of the Shapiro resonance of Ring 1, obtained at an applied frequency of 144 MHz and plotted as a histogram map}\label{Figure1}
 \end{figure*}

Strontium ruthenate (Sr$_2$RuO$_4$) is one of the most extensively studied TRSB superconductors, with compelling evidence from zero-field $\mu$SR~\cite{Luke2000,Grinenko2021,grinenko__2023} and Kerr rotation experiments~\cite{Xia2006} on bulk \textit{macroscopic} crystals. Yet, attempts to observe the TRSB on a \textit{microscopic} scale, e.g. via scanning-SQUID, have been inconclusive~\cite{Hicks2010,Curran2014,curran_search_2023}. Consequently, the microscopic origin of TRSB in Sr$_2$RuO$_4$ and its connection to the pairing symmetry remain unresolved~\cite{Maeno2012,Mackenzie2017,Junctions2021,maeno_still_2024}.

There have been a series of attempts to verify the violation of $I_\text{c+}(H) = |I_\text{c-}(-H)|$ in Sr$_2$RuO$_4$-hybrid proximity Josephson junctions, where conventional superconductors (e.g., Nb) are connected through a bulk Sr$_2$RuO$_4$ crystal~\cite{Kidwingira2006,Anwar2013,Anwar2017,TRS_not_broken_helical}. Such devices exhibit numerous unusual transport characteristics, including hysteretic $I_\text{c}(H)$ patterns --- which alter their shape when the sample is thermally cycled above $T_\text{c}$ --- and stochastic switching of critical current over time (i.e., telegraph noise) as a function of applied magnetic field and bias current. These features have been accounted for by the emergence of finite-sized superconducting domains with an arbitrary configuration. Here, each domain corresponds to a degenerate state ($\Psi^+$ and $\Psi^-$) of the two-component order parameter. Alternatively, it highlights the possible multi-band pairing in Sr$_2$RuO$_4$, which results in supercurrent rectification and TRSB, when interfaces with a conventional superconductor~\cite{yerin_supercurrent_2024,yerin_supercurrent_2025,ng_broken_2009}. Note, the two-component nature of the order parameter was later corroborated by the ultrasound experiments~\cite{Ghosh2020,Benhabib2020,ghosh_strong_2022}. The hybrid-Sr$_2$RuO$_4$ proximity devices are commonly characterized by irregular switches in their supercurrent interference pattern, which can indicate current or magnetic field-driven domain wall motion. However, these stochastic irregularities make it impossible to verify the violation of $I_\text{c+}(H) = |I_\text{c-}(-H)|$. This requires a continuous and reproducible interference pattern, interdependent of magnetic field sweep direction, which can not be realized in macroscopic crystals used in previous studies on hybrid-Sr$_2$RuO$_4$ junctions.

Nanostructured Sr$_2$RuO$_4$ crystals, on the other hand, offer a distinct possibility to gain control over the proposed superconducting domain structure, without the need for the proximity effect through superconducting contacts. In analogy to ferromagnetism, geometrical constrictions can serve as the most energetically favorable sites for domain wall formation and pinning. This enables stabilizing an isolated domain wall and studying it in detail through repeated phase-sensitive transport experiments. A domain wall is a finite region within the condensate, where the degenerate states of the order parameter smoothly transition into another, resulting in a local suppression of the condensate, resulting in an intrinsic Josephson coupling, as was first proposed by Sigrist and Agterberg~\cite{sigrist_role_1999}. Essentially, if the superconducting order parameter of Sr$_2$RuO$_4$ can host superconducting domains, confinement allows us to study the degenerate states from neighboring domains that partially overlap at a DW through their Josephson coupling. See Figure \ref{Figure1}a for a conceptual depiction of a constricted device that captures a superconducting domain wall.

Here we return to the Josephson coupling in single-crystal Sr$_2$RuO$_4$ microstructures where Josephson junctions arise \textit{intrinsically} in the superconducting phase: in a previous work, we found a two-channel critical current interference pattern in single-crystal rings of Sr$_2$RuO$_4$, which could not be explained by conventional constriction junctions~\cite{Yasui2020}. Rather, we showed that these superconducting quantum interference device (SQUID) oscillations matched predictions by time-dependent Ginzburg-Landau simulations of a two-component superconductor with a symmetry-protected chiral pairing (e.g., chiral d-wave), which features the formation of stable domain walls across the ring; relating the Josephson coupling to the local suppression of the condensate at the DW (see Figure \ref{Figure1}a). We also showed that the Josephson effect only appears in samples exhibiting an intrinsic $T_\text{c}$ of unstrained Sr$_2$RuO$_4$ ($T_\text{c}\sim1.5~\text{K}$), and is completely absent in those in the extrinsic \textit{3K-phase}: a phase with increased critical temperature ($T_\text{c}\sim3~\text{K}$) induced by strain~\cite{ying_enhanced_2013,Steppke2017}. Despite their similar dimensions and material quality (e.g., residual resistivity), the mesoscopic rings with a \textit{3K-phase} do not show SQUID oscillations~\cite{Yasui2017,Yasui2020}. This is consistent with the recent studies of the \textit{3K-phase}, which demonstrate that the \textit{3K-phase} is dominated by a single-component order parameter~\cite{Anwar2017,Grinenko2021}.

In this work, we show that these spontaneous junctions break time-reversal symmetry by violating $I_\text{c+}(H) = |I_\text{c-}(-H)|$, as well as display a wealth of unconventional transport signatures: a diode effect at zero applied magnetic field, interference patterns with anomalous periods deviating from the flux quantum $\Phi_0 = h/2e$, a complete breakdown of the RCSJ model, fractional Shapiro steps, and current-switchable bistable states with strongly nonreciprocal transport. Our data is consistent with theoretical predictions for the unusual Josephson potential of a chiral domain wall, which can be viewed as a hybrid of $\varphi_0$- and $\varphi$-junctions~\cite{sigrist_role_1999}.

\section*{Results}

\subsection*{Unconventional Josephson junctions in microstructured $\mathrm{Sr}_2\mathrm{RuO}_4$}

Using the sample fabrication process described in the Methods section, we have produced various singly and doubly connected Sr$_2$RuO$_4$ microstructures, contacted with non-superconducting leads. Therefore, all superconducting correlations stem from the Sr$_2$RuO$_4$ crystals. Our devices are characterized by high ($>$ 100) residual resistivity ratios and a sharp superconducting transition below 1.5 K, showing the pristine quality of the samples and absence of the extrinsic 3K-phase, strain or any edge dislocations~\cite{ying_enhanced_2013,Steppke2017}. The data presented here is obtained on three rings, but the features discussed here are universally observed in a large selection of samples. See Figure \ref{Figure1}b for a false-colored scanning electron micrograph of Ring 1. The rings have a typical inner diameter of $\sim500$ nm, and a typical outer diameter of 1.2 \textmu m, compared to the zero temperature ab-plane superconducting parameters of Sr$_2$RuO$_4$: $\xi_0 =$ 66 nm and $\lambda=$ 190 nm \cite{maeno_still_2024}. For an extended dataset including all spatial dimensions, the reader is referred to Supplementary Note S1.

 \begin{figure*}[h!]
 \centerline{$
 \begin{array}{c}
 \includegraphics[width=0.77\linewidth]{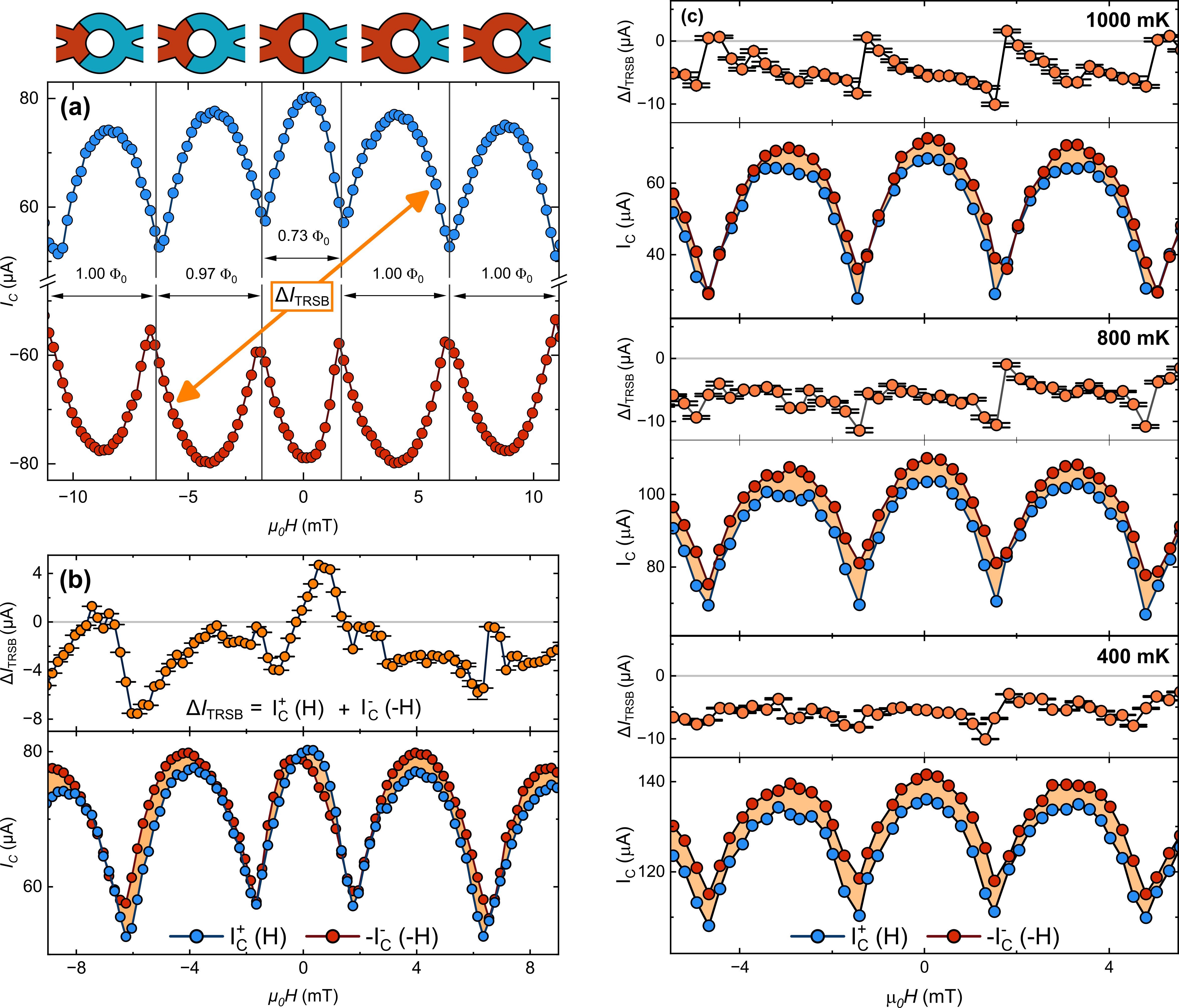}
 \end{array}$}
 \caption{\textbf{Time-reversal symmetry breaking (TRSB) in the critical current interference ($I_\text{c}(H)$-)pattern of microscopic ring samples.} (a) Superconducting Quantum Interference Device (SQUID) oscillations as recorded on Ring 1 at 50 mK. The central lobe is skewed in the same direction for both current polarities. Each lobe width is expressed in magnetic flux quantum ($\Phi_0 = \frac{h}{2e}$), according to the ring area ($\pi r_{\text{in}}r_{\text{out}}$, where $r_{\text{in}}$ and $r_{\text{out}}$ are the inner and outer ring diameter respectively). The center lobe is noticeably smaller than $\Phi_0$, and is skewed in the same direction for both current polarities. The drawings illustrate the toy model, accounting for the period anomaly, based on field-driven domain wall displacement. (b) The $I_\text{c}(H)$ data shown in (a), where $I_\text{c-}(-H)$ superimposed on $I_\text{c+}(H)$. The TRSB is accentuated by the orange-shaded area between the curves in the bottom panel, and is quantified in the top panel as $\Delta I_{\text{TRSB}}=I_\text{c+}(H) - |I_\text{c-}(-H)|$. Note, this does not represent the superconducting diode effect, since we superimpose $I_\text{c-}(-H)$ on $I_\text{c+}(H)$, this constitutes essentially as a check for point symmetry in the origin. (c) TRSB in Ring 2, measured at three different temperatures (1000 mK, 800 mK, and 400 mK respectively from top to bottom). Note that $I_\text{c-}(0) \neq I_\text{c+}(0)$ indicating a magnetic field-free Josephson diode effect.}\label{Figure2}
 \end{figure*}

All devices show a robust Josephson effect as evidenced by the clear appearance of Shapiro steps obtained in the $IV$-characteristic, as shown in Figure \ref{Figure1}c. Besides, in Figure \ref{Figure1}d we show the evolution of the Shapiro steps as a function of applied microwave power. Note, since Sr$_2$RuO$_4$ is the only superconductor in the samples, the observed Josephson effect must originate from spontaneously appearing junctions in the rings. The Shapiro steps are present in a variety of geometries, including devices in a singly-connected configuration (see Supplementary Figure S3).

Next, we study the TRSB of the observed Josephson effect through the supercurrent interference pattern. Figure \ref{Figure2}a shows the $I_\text{c}(H)$-pattern recorded at 50 mK on Ring 1, which is exemplary for all our ring samples, showing clear $I_\text{c}$ oscillations. Since Shapiro steps can only originate from the Josephson effect, we can rule out Little-Parks oscillations or kinematic vortices as origin of the oscillatory $I_\text{c}(H)$-pattern. Rather, this implies that two Josephson junctions spontaneously appear in both arms of the ring, once the crystal is cooled below $T_\text{c}$. This implies that the resistance oscillations observed in similar devices measured by Cai et al.~\cite{cai_unconventional_2013,cai_magnetoresistance_2022}. should be interpreted as resulting from critical current oscillations associated with Josephson coupling rather than $T_\text{c}$ oscillations originating from the Little-Parks effect.

\subsection*{Spontaneous TRSB of the Josephson junctions}

We probe the breaking of time-reversal symmetry using two main methods. First, we examine the zero-field diode effect, which requires TRSB. Secondly, time-reversal symmetry dictates that the critical current remains unchanged when we change both the current polarity and the magnetic field direction, which we express as $I_\text{c+}(H) = |I_\text{c-}(-H)|$, essentially examining point symmetry in the origin. In this section, we show that our devices violate these simple rule --- supporting TRSB in the intrinsic phase of Sr$_2$RuO$_4$ ($T_\text{c}$ $\sim$ 1.5 K) --- and highlight other unconventional features in the $I_\text{c}(H)$-patterns.

In Figure \ref{Figure2}, we present the analysis of the interference patterns of Ring 1 and Ring 2. We account for all the systematic measurement artifacts and quantify the uncertainty arising from stochastic effects (such as IV curve rounding, voltage offsets, long-term drifts, and remnant fields in the superconducting magnet). For the details of this analysis method, see Supplementary Note S2.

The breaking of time-reversal symmetry is visualized by Figure \ref{Figure2}b and Figure \ref{Figure2}c, which superimposes $I_\text{c-}(-H)$ on $I_\text{c+}(H)$, by flipping the negative $I_\text{c}(H)$-branch in both the $H$-axis and $I$-axis, as demonstrated by the orange arrow in Figure \ref{Figure2}a. We shade the difference between $I_\text{c+}(H)$ and $|I_\text{c-}(-H)|$ in orange. In order to highlight the difference between and $I_\text{c+}(H)$ and $|I_\text{c-}(-H)|$, we plot $I_\text{c+}(H)\:- |I_\text{c-}(-H)|$, yielding the parameter:

\begin{align}\label{eq1_1}
\Delta I_{\text{TRSB}}  =I_\text{c+}(H) - |I_\text{c-}(-H)|
\end{align}

Here, a finite $\Delta I_{\text{TRSB}}$ signifies a finite TRSB, which we have observed in multiple devices and across multiple experimental setups. Note that this relation is not to be confused with the generalized superconducting diode effect, $\Delta I_{\text{diode}} =I_\text{c+}(H) - |I_\text{c-}(H)|$, except for $H=0$, where $\Delta I_{\text{TRSB}}(0) = \Delta I_{\text{diode}}(0)$, i.e., the zero-field diode effect. In Figure \ref{Figure2}c, we show the temperature evolution of the effect in Ring 2. We also provide an additional example of TRSB in the Supplementary Note S1. All devices exhibit a zero-field superconducting diode effect and a violation of $I_\text{c+}(H) = |I_\text{c-}(-H)|$, which exceeds the statistical measurement error of our experiments.

In addition to the violation of $I_\text{c+}(H) = |I_\text{c-}(-H)|$ and the field-free Josephson diode effect, we observe an anomalous period of the SQUID oscillations. As illustrated in Figure \ref{Figure2}a, the central lobe of the $I_\text{c}(H)$-pattern is significantly narrower (0.73 $\Phi_0$) than the neighboring lobes, which is observed independently of field sweep direction and appears for all three rings presented here. This peculiar departure from $\Phi_0$-periodicity cannot be reconciled with the established magnetotransport mechanisms for SQUIDs. We note that multi-band proximity SQUIDs can feature anomalous patterns that deviate from the standard RCSJ SQUID~\cite{yerin_josephson_2015}, although such patterns remain $\Phi_0$-periodic, and are therefore not consistent with our findings. Qualitatively, our results appear to be most consistent with the theoretical predictions of field-induced domain wall motion. This scenario is illustrated in the drawings above Figure \ref{Figure2}a, and is further detailed in the discussion.

Besides its width, the central lobe of the $I_\text{c}(H)$-pattern is skewed, while the surrounding lobes are not. For opposite current polarities, the central lobes are skewed in the same field direction, rather than in opposite field directions. Although skewed lobes can occur in regular SQUIDs as a result of a geometrical asymmetry, all lobes would skew in equal amounts, which is not the case for Ring 1. Besides, the highly symmetric design of our devices makes a geometric asymmetry as the underlying origin of the skewed $I_\text{c}(H)$-pattern highly unlikely.

In conclusion, the interference patterns obtained on our Sr$_2$RuO$_4$ devices cannot be captured by the conventional RCSJ model. In this model, device asymmetries such as mismatched critical currents or electrode lengths lead to asymmetric interference patterns, but the identity $I_\text{c+}(H) = |I_\text{c-}(-H)|$ is always preserved. Any deviation from this identity substantiates evidence for{~\cite{CIT_Cuozzo,trimble_josephson_2021,TRS_not_broken_helical, wu_field_free_2022, szombati_josephson_2016, baumgartner_supercurrent_2022}. To illustrate this, we present a typical fit of the RCSJ model to a conventional SQUID in Supplementary Note S4 and show that it preserves time-reversal symmetry. Besides, in the RCSJ model, geometric asymmetries generate skewed lobes; however, they would never break time-reversal symmetry or generate a zero-field superconducting diode effect. Besides, it would skew every lobe in the pattern equally.

\subsection*{Manipulating unconventional junctions using current and ab-plane magnetic fields}

By applying small in-plane magnetic fields along the bias current direction --- which we define as the x-direction (see inset of Figure \ref{Figure3}a) --- we reveal a striking bistable $I_\text{c}$ state, which can be controllably switched by application of a training bias current. We coin these states the high-$I_\text{c}$ and low-$I_\text{c}$-state. In Figure \ref{Figure3}a we plot these two $I_\text{c}$-branches that appear under the application of $\mu_0H_{\text{x}}$ = 20 mT: a high $I_\text{c}$ (blue curve) and a low $I_\text{c}$ (orange curve), along with a reference curve taken at zero field (pink curve). By applying a training current bias above a certain current threshold, we can switch between these states. For example, if the sample is initialized in the low $I_\text{c}$ configuration, we find a consistently low critical and retrapping current on both current polarities, lacking any hysteresis (Figure \ref{Figure3}b). However, when we apply a large enough positive current, the device switches to the high $I_\text{c}$-branch, which results in an apparent hysteresis loop observed on the positive current polarity (Figure \ref{Figure3}c).

 \begin{figure*}[h!]
 \centerline{$
 \begin{array}{c}
 \includegraphics[width=1\linewidth]{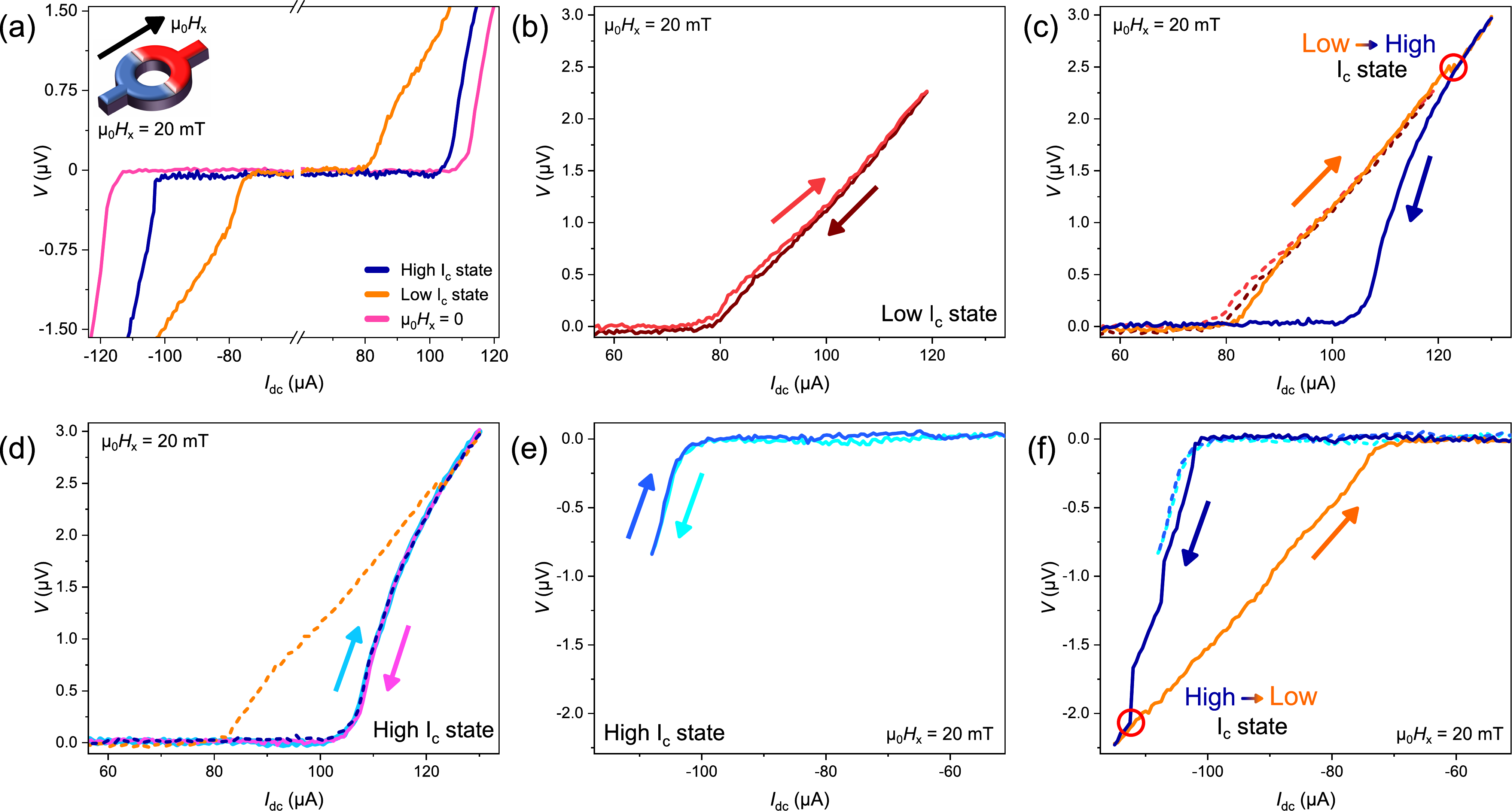}
 \end{array}$}
 \caption{\textbf{Two-level critical current state revealed under the application of an in-plane magnetic field.} (a) Comparison between the zero-field current-voltage ($IV$)-characteristic (pink curve) and the two stable critical current ($I_\text{c}$)-branches that appear under the application of an in-plane magnetic field of 20 mT along the $x$-direction (see inset of (a)): the blue curve is the high $I_\text{c}$ state, and in orange, we plot the low $I_\text{c}$ state. (b) The low $I_\text{c}$ state is stable for a large range of currents and no hysteresis is observed when sweeping the current. (c) If the bias current is increased above a threshold training value, the device switches from the low $I_\text{c}$ state to the high $I_\text{c}$ state (here, the training current is $122$~{\textmu}A). (d) and (e) the high $I_\text{c}$ state is subsequently stable on both current polarities. (f) The reverse operation is performed by applying a bias current above a negative threshold value (here, the training current is $-114$~{\textmu}A). All $IV$-characteristics are obtained on Ring 3 at 600 mK. In all panels except (a), the arrows denote sweep direction.}\label{Figure3}
 \end{figure*}

Now, the sample is stable in the high $I_\text{c}$-state (i.e., a high $I_\text{c}$, which equals the retrapping current; Figure \ref{Figure3}d). The high $I_\text{c}$-branch is stable for both current polarities (Figure \ref{Figure3}e). Similarly, when a large enough negative training bias current is applied the sample switches back to the low critical current state, resulting in an apparent hysteresis loop on the negative current polarity (Figure \ref{Figure3}f).

Note that both $I_\text{c}$ states are exceedingly stable: the apparent hysteresis loops only occur during a transition between the $I_\text{c}$ states. A hysteresis loop is never observed while cycling the current multiple times on the same polarity. Instead, the switching loops alternate between current polarities, making this behavior completely different from hysteresis loops in underdamped Josephson junctions.

We find that field cooling or applying the field after cooling the sample below $T_\text{c}$ gives identical results. Besides, we find that the relation between training current polarity and resulting high or low $I_\text{c}$-branches is exchanged when the magnetic field direction is reversed. In other words, when reversing the field direction, a positive training current results in the low $I_\text{c}$ state (as opposed to the high $I_\text{c}$ state as shown in Figure \ref{Figure3}c). To illustrate this, we present the $IV$-characteristics for $\mu_0H_{\text{x}}$ = -20 mT in Supplementary Note S6. Finally, we observe this bistable $I_\text{c}$ property in all samples, irrespective of device geometry. On the other hand, we find that the direction and magnitude of the in-plane magnetic field, required to obtain this behavior, varies between samples. For a more detailed discussion, the reader is referred to Supplementary Note S6.

\subsection*{Exploring the link between switchable $I_\text{c}$-states and the current-phase relationship}

To further highlight the unconventional nature of the bistable $I_\text{c}$ state, we examine the current-phase relation using the AC-Josephson effect under the application of in-plane fields. We found that some combinations of field magnitude and alignment can give rise to clear half-integer Shapiro steps, indicating a deviation from the standard sinusoidal current-phase relation. 

In Figure \ref{Figure4}a, we plot both low and high $I_\text{c}$ state of Ring 1, induced by the application of a 10 mT magnetic field along the $x$-direction, with simultaneous irradiation of microwaves. We observe that the high $I_\text{c}$-branch only shows an integer Shapiro response, whereas the low $I_\text{c}$-branch shows both integer and half-integer steps. This suggests that the current-phase relation differs between the two $I_\text{c}$ states.

 \begin{figure}[h!]
 \centerline{$
 \begin{array}{c}
 \includegraphics[width=0.6\linewidth]{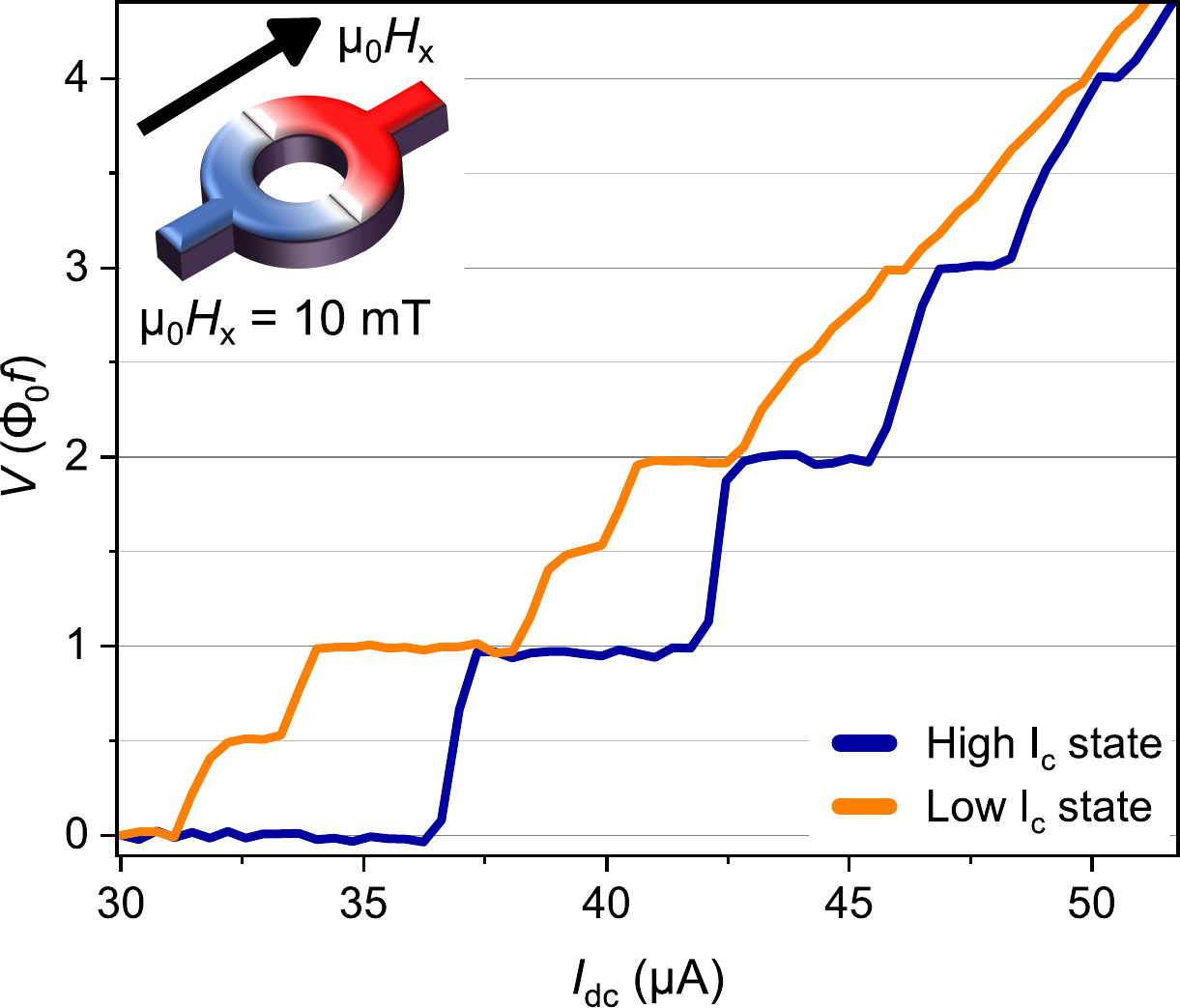}
 \end{array}$}
 \caption{\textbf{Bistable current-voltage ($IV$)-characteristics obtained under the simultaneous irradiation of microwaves and application of in-plane magnetic field.} The low critical current ($I_\text{c}$)-branch shows both integer and half-integer steps, whereas the high $I_\text{c}$ state only shows integer steps. This indicates a different current-phase relationship between the two $I_\text{c}$ states. All data obtained on Ring 1.}\label{Figure4}
 \end{figure}

Note that the half-integer steps cannot be explained by the flow of in-plane Abrikosov vortices, as the Josephson penetration depth is far larger than the width of the Josephson junction (i.e., set by the dimensions of the arm of the ring)~\cite{Sellier2004,flux_flow}. 

Finally, we find that the Shapiro response becomes highly asymmetric under the application of in-plane fields: the half-integer steps are limited only to a single current polarity (in this case the positive polarity). We found that reversing the field direction changes the bias polarity showing the fractional steps, indicating that there is a relation between the bistable $I_\text{c}$ states and the appearance of the fractional Shapiro steps: both properties require in-plane magnetic fields of similar magnitude. Finally, there seems to be no correspondence between the geometry of the sample and the field direction required to trigger either the two-level $I_\text{c}$ states or the non-integer Shapiro steps.

\section*{Discussion}

Our devices spontaneously break time-reversal symmetry, show interference patterns with periods deviating from the flux quantum, are not compatible with the RCSJ model, exhibit fractional Shapiro steps, and current-switchable bistable states with strongly nonreciprocal transport. These features cannot originate from conventional s-wave superconductivity. Rather, we find that our data is consistent with superconducting domain wall junctions and a two-component order parameter.

A two-component order parameter featuring superconducting domains naturally results in a superconducting state that features both TRSB and spontaneous Josephson coupling. For Sr$_2$RuO$_4$ specifically, this two-component order parameter picture is supported by ultrasound measurements~\cite{Ghosh2020,Benhabib2020}. When it comes to domain superconductivity, more recent ultrasound experiments seem to confirm the presence of mobile superconducting domains~\cite{ghosh_strong_2022} and experiments on Nb-Sr$_2$RuO$_4$ hybrid-junctions revealed distinct telegraph noise near the critical current, which was interpreted as originating from moving domain walls~\cite{Kidwingira2006,Anwar2013,Anwar2017}. Finally, note that similar transport experiments as presented here, conducted on the Kagome superconductors, were interpreted as resulting from domain superconductivity~\cite{Le2025,lou_radio_frequency_2025,blom_emergent_2026,rog_direct_2026}.

Continuing on the breakdown of the RCSJ model: if the domains originate from chiral states with opposite orbital phase windings~\cite{Yasui2020}, each domain contributes a finite magnetic flux to the loop. However, the two chiral states generate flux in opposite directions. At zero external field, where the domain wall is centered, the contributions of the domains cancel out. However, if we apply an external field in either direction, one chirality is promoted over the other --- resulting in a shift in the domain wall position (illustrated in the drawings above Figure \ref{Figure2}a). This enhances the coupling of one chirality to the loop, adding an additional magnetic flux to the applied field.

Due to this enhancement, the total flux through the loop reaches $\Phi_0$ at a smaller applied field than it would for a regular SQUID, explaining the smaller middle lobe. However, the field-induced DW motion is limited to a finite range within the arms of the ring, determined by the constricted necks on each side of the loop. Therefore, the pattern ultimately regains its $\Phi_0$ periodicity. In Supplementary Note S3, we apply this toy model with a phenomenological parametrization of field-induced domain wall displacement in attempt to fit multiple interference patterns. Furthermore, in Supporting Figure S5, we present an example of a doubly-connected structure without constricted leads to halt the field-driven DW motion. The interference pattern of this device becomes noticeably more irregular and noisy.

The bistable $I_\text{c}$ property reported here can also be explained by the presence of superconducting domains in Sr$_2$RuO$_4$. In experiments on macroscopic Nb-Sr$_2$RuO$_4$ hybrid-junctions, the telegraph noise was found to increase with the application of small ab-plane magnetic fields, suggesting that the domain walls can be moved under the application of ab-plane magnetic fields and bias currents~\cite{Kidwingira2006}. Moreover, in Nb-Sr$_2$RuO$_4$ hybrid-junctions hysteresis loops are observed that bear remarkable similarity to the ones presented here~\cite{Anwar2013}. In conclusion, this indicates that the domain walls become less strongly pinned due to ab-plane magnetic fields.

In contrast to previous experiments, the domain walls are more strongly pinned in the microstructures presented here. In the absence of magnetic fields, the domain walls are deterministically pinned. However, when ab-plane magnetic fields are applied, the pinning becomes weakened. Since a different domain configuration can lead to a different $I_\text{c}$, a bistable $I_\text{c}$ state implies a bistable domain configuration~\cite{Bouhon2010}. Therefore, we assert that the telegraph noise associated with uncontrolled domain wall movement observed in bulk samples is of the same origin as the bistable critical current in our microscopic samples, namely the underlying domain configuration.

The half-integer Shapiro step observed in the $I_\text{c}$ state can also support the domain wall hypothesis. Sigrist and Agterberg showed that the Josephson energy of a superconducting chiral DW has two minima at $\varphi_0$ and $\varphi$~\cite{sigrist_role_1999}. Interestingly, depending on the orientation of the DW, the two states can change their relative phase, resulting in an anomalous phase of the Josephson junction. For example, when the transport current is normal to the DW, the ground state corresponds to a regular 0-junction, whereas a DW angle of 45 degrees, results in a $\varphi$-junction. This can cause different domain wall states, to have different current-phase relationships. We discuss this further in Supplementary Note S5. Additionally, the movement of superconducting domain walls can also explain the relatively small width of the first lobe in the $I_\text{c}(H)$-pattern compared to the neighboring lobes shown in Figure \ref{Figure2}a. See Supplementary Note S3 for a model that captures this behavior in terms of domain wall movement. On the other hand, how current and ab-plane magnetic fields couple to the domain wall configuration remains an open question.

Finally, it must be noted that similar in-plane magnetic fields were used in the magnetometry experiments that showed evidence of half-quantum vortices in mesoscopic Sr$_2$RuO$_4$ rings~\cite{Jang2011}. This calls for a re-examination of these results and a possible reinterpretation based on superconducting domain walls present in these rings.

We now turn to the nature of the two-component order parameter in Sr$_2$RuO$_4$. Our experiments suggest a two-component order parameter that naturally breaks into spatially segregated domains that spontaneously stabilize below $T_c$, resulting in a TRSB state. A symmetry-protected chiral pairing naturally satisfies these criteria. This would, for example, favor a pairing in the $E_g$ irreducible representation of $D_{4h}$, i.e., $d_{xz} + id_{yz}$. It must be noted, however, that a $d_{xz} + id_{yz}$ pairing is expected to show a sizable second discontinuity in the heat capacity and the condensate density when entering the TRSB state, which is not detected in experiments~\cite{Li2021,mueller_constraints_2023}. 

Alternatively, order parameters composed out of an accidental degeneracy between two single-component order parameters can split into spatially segregated domains~\cite{Kivelson2020,Romer2021,scaffidi_degeneracy_2023}. However, this requires a varying strain texture caused by edge-dislocations and impurities for the stabilization of the domain walls and a TRSB state~\cite{Yuan2021,Willa2021}. This is contradictory to our ultra-pure and strain-free crystals. Besides, this picture suggests that the domains are deterministically pinned by the crystal lattice and therefore are not mobile, which is opposed by our experiments, and previous work observing dynamic domains~\cite{Kidwingira2006,Anwar2013,Anwar2017}. Finally, based on elastocaloric measurements, order parameters that rely on two nearly degenerate competing parings are unlikely~\cite{li_elastocaloric_2022,palle_constraints_2023}.

\section*{Conclusion}

In conclusion, we have shown spontaneous TRSB Josephson coupling in ultra-clean Sr$_2$RuO$_4$ microstructures supported by the measurement of the $I_\text{c}(H)$-pattern, which violates $I_\text{c+}(H) = |I_\text{c-}(-H)|$, and includes a magnetic field-free Josephson diode effect. This is accompanied by a wealth of unconventional features: under the application of in-plane fields, a bistable critical current emerges, characterized by a low $I_\text{c}$ and a high $I_\text{c}$ state that can be reliably switched by a training bias current. Besides, we have found that the current-phase relation of these bistable critical current states can be altered, which is reflected by the appearance of half-integer steps in the Shapiro response. 

The combination of these unusual junction properties cannot originate from an s-wave superconductor but is consistent with domain wall superconductivity, associated with a multi-component order parameter. In particular, harnessing these domain walls in functional devices point towards an independent route to field-free Josephson diodes, memory elements, and phase batteries, which so far could only be realized in complex proximity weak links. Superconducting domain walls are a defining feature of multi-component order parameter superconductors, and their Josephson transport provides a rich platform to examine the pairing symmetry of Sr$_2$RuO$_4$ and multi-component order parameter systems in general.

\section*{Methods}

High-quality Sr$_2$RuO$_4$ single crystals are grown using the floating zone method~\cite{Mao2000}. They are subsequently transferred on highly resistive SrTiO$_3$ substrates using mechanical exfoliation. This results in flat flake-like microscopic crystals, which are significantly thicker than monolayers; typically 200 nm thick. Therefore, random strain patterns, wrinkles, and folds, normally associated with the thin film limit, are absent. Note that the c-axis of the flakes is perpendicular to the substrate plane. Next, the crystals are contacted using sputter-deposited Ti/Au contacts in an electron beam lithography lift-off process. They are capped using a SiO$_\text{x}$ or gold capping layer. Finally, we use focused Ga-ion beam (FIB) milling to structure the crystals in micron-sized devices. The combination of ultra-low beam currents and the capping layer is essential to prevent damage to the crystals by the high-energy Ga-beams. The ultra-low beam currents ensure that any structural damage is highly localized (limited to 10 nm from the side edges of the milled structure by TEM analysis~\cite{cai_unconventional_2013}), and the capping layer protects the top of the crystal during the inevitable radiation of Ga-ions. This process yields devices with high ($>$ 100) residual resistivity ratios while retaining bulk $T_\text{c}$ values, even after remilling the crystal several times (see Supporting Figure S2). Neither the exfoliation nor the focused ion beam structuring reduces the quality of the crystal. The $T_\text{c}$ of our devices never exceeds 1.5 K, showing the absence of the extrinsic \textit{3K-phase} and demonstrating the absence of any edge-dislocations or strain in our devices~\cite{ying_enhanced_2013,Steppke2017}. 

\section*{Author contributions}

KL and JA conceived the project and coordinated the research. YM provided the crystals. RF carried out the transport experiments with help from GS, DW, CL, and JR. MR carried out the TRSB analysis and provided the phenomenological model. RF, MR and KL wrote the manuscript with inputs from all authors.

\section*{Funding information}

This work was financed by the Dutch Research Council (NWO) through the "Enhancing superconducting diodes with spin-polarized supercurrents" (project 019.231EN.005),  “The Full Picture: Novel Microscopy with Smart Quantum Probes” (project VI.Veni.212.302), “Three seeds for the quantum/nano-revolution” (Project NWA 1418.22.001), and the Frontiers of Nanoscience (NanoFront) program. It was also supported by EU Cost Action CA16218 (NANOCOHYBRI).

\section*{Competing interests}

The authors declare no competing interests.

\section*{Data availability}

The data supporting this study are openly available in the repository 4TU.ResearchData at https://doi.org/10.4121/56863876-5299-44a0-bb24-75540b3ba797.

\section*{Acknowledgments}

We are grateful to Alexander Brinkman (University of Twente) for helpful discussions. This project benefited from access to The Netherlands Centre for Electron Nanoscopy (NeCEN) at Leiden University.


\end{document}


\title{Supplemental Material of: Time-reversal symmetry breaking and inherent Josephson diodes in a multi-component order parameter superconductor}

\date{\today}

\author{Remko Fermin}
\affiliation{Department of Materials Science \& Metallurgy, University of Cambridge, 27 Charles Babbage Road, Cambridge CB3 0FS, United Kingdom}
\affiliation{MESA+ Institute for Nanotechnology, University of Twente, 7500 AE Enschede, The Netherlands}
\author{Matthijs Rog}
\author{Guido Stam}
\affiliation{Huygens-Kamerlingh Onnes Laboratory, Leiden University, P.O. Box 9504, 2300 RA Leiden, The Netherlands}
\author{Daan Wielens}
\author{Joost Ridderbos}
\author{Chuan Li}
\affiliation{MESA+ Institute for Nanotechnology, University of Twente, 7500 AE Enschede, The Netherlands}
\author{Yoshiteru Maeno}
\affiliation{Toyota Riken – Kyoto University Research Center (TRiKUC), Kyoto 606-8501, Japan}
\author{Jan Aarts}
\author{Kaveh Lahabi}
\email{lahabi@physics.leidenuniv.nl}
\affiliation{Huygens-Kamerlingh Onnes Laboratory, Leiden University, P.O. Box 9504, 2300 RA Leiden, The Netherlands}

\pacs{} \maketitle

\clearpage

\section{Extended data}

\begin{figure}[h!]
 \centerline{$
 \begin{array}{c}
 \includegraphics[width=0.9\linewidth]{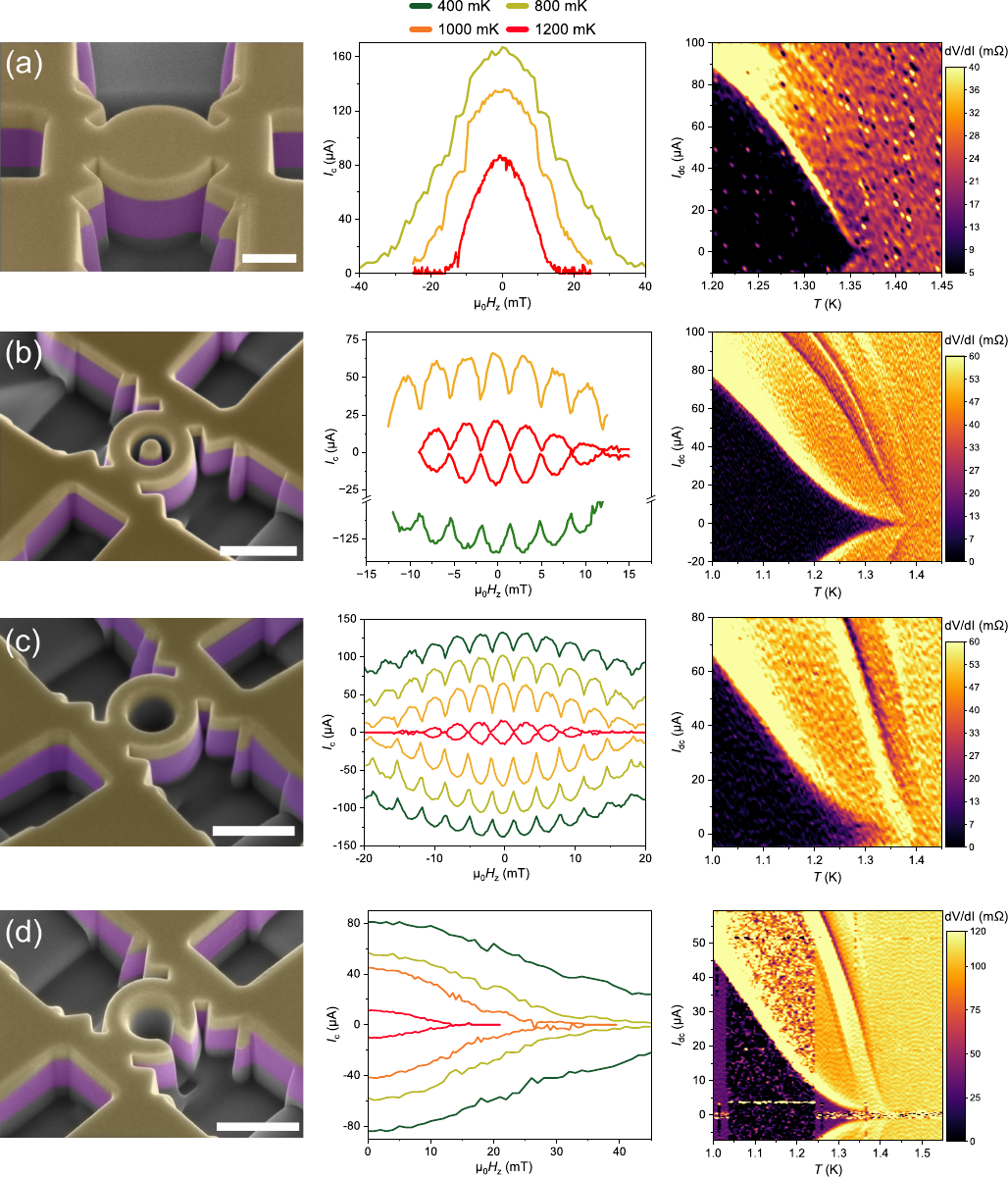}
 \end{array}$}
 \caption{Full dataset obtained on Ring 2. Row (a) shows a false-colored SEM micrograph of the initial geometry of the sample: a disk. The Sr$_2$RuO$_4$ crystal is colored purple and the gold protective layer is yellow. Note the dark contrast on top of the gold, caused by an additional protective layer of SiOx. The $I_{\text{c}}(B)$-pattern at various temperatures and a colormap of d$V$/d$I$ versus current and temperature accompany the SEM micrograph to the right. Likewise for subsequent sample geometries depicted in rows (b), a ring with a centered island, (c), a ring, and (d), a ring with one of the arms milled away. The scale bar corresponds to 1~{\textmu}m in all SEM micrographs.}\label{C15_table}
 \end{figure}

In this paper, we provide data obtained on four different devices. Although the dimensions of these devices differ slightly, all measured devices show signatures of domain wall junctions, including unconventional properties like the bistable critical current property. An overview of the samples presented in the main text is shown in Figure \ref{table2}. On Ring 2 we have carried out four steps of FIB processing and subsequent measurements. Figure \ref{C15_table} shows an overview of the $I_{\text{c}}(B)$-patterns and the critical current as a function of the temperature of this device.

 \begin{table}[h!]
 \centerline{$
 \begin{array}{c}
 \includegraphics[width=0.7\linewidth]{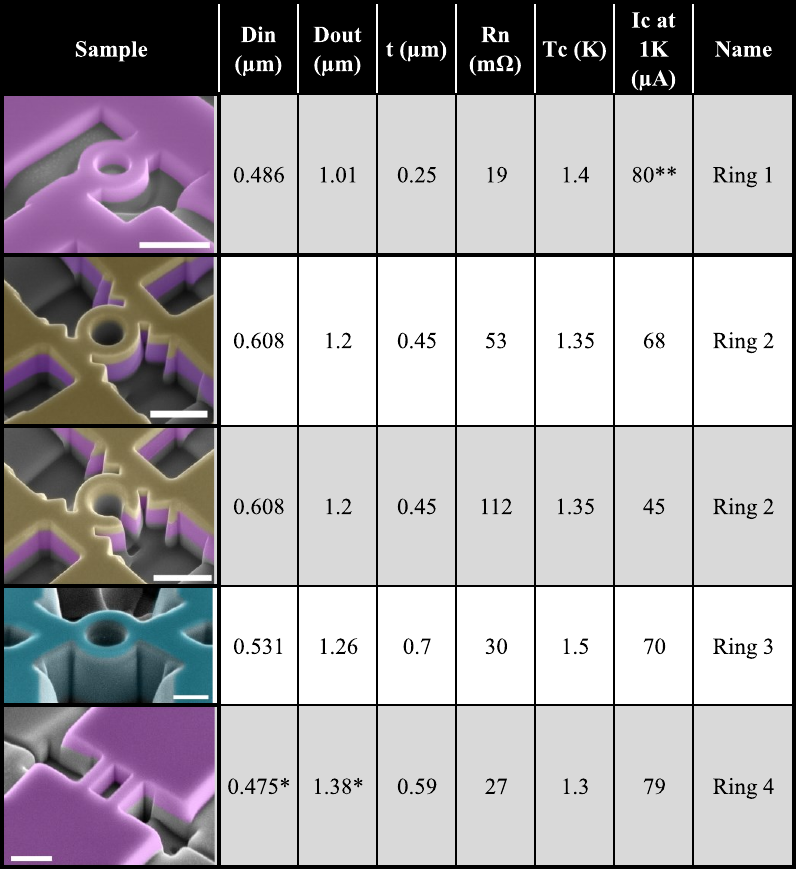}
 \end{array}$}
 \caption{Table containing the dimensions (inner diameter $D_{\text{in}}$, outer diameter $D_{\text{out}}$ and the thickness $t$) and other details of the ring samples used acquire the data presented in this paper.}\label{table2}
 \end{table}

 \clearpage

\begin{figure}[h!]
 \centerline{$
 \begin{array}{c}
 \includegraphics[width=0.6\linewidth]{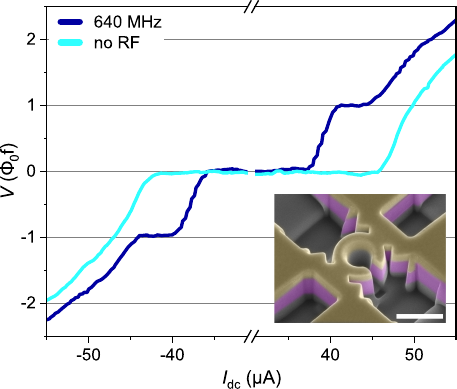}
 \end{array}$}
 \caption{Shapiro response of the single arm of Ring 2 obtained at 1~K. The voltage is normalized to the expected height of the Shapiro steps at 640 MHz. The Shapiro steps confirm the presence of a chiral domain wall after cutting one arm of the ring and transforming it into a singly-connected sample.}\label{single_shapiro}
 \end{figure}

Specifically interesting is the final state of Ring 2, in which we have cut one arm of the ring to form a singly connected sample. First, we confirmed the Josephson coupling by obtaining Shapiro steps on this singly-connected geometry, as shown in Figure \ref{single_shapiro}. 

\begin{figure}[t!]
 \centerline{$
 \begin{array}{c}
 \includegraphics[width=0.8\linewidth]{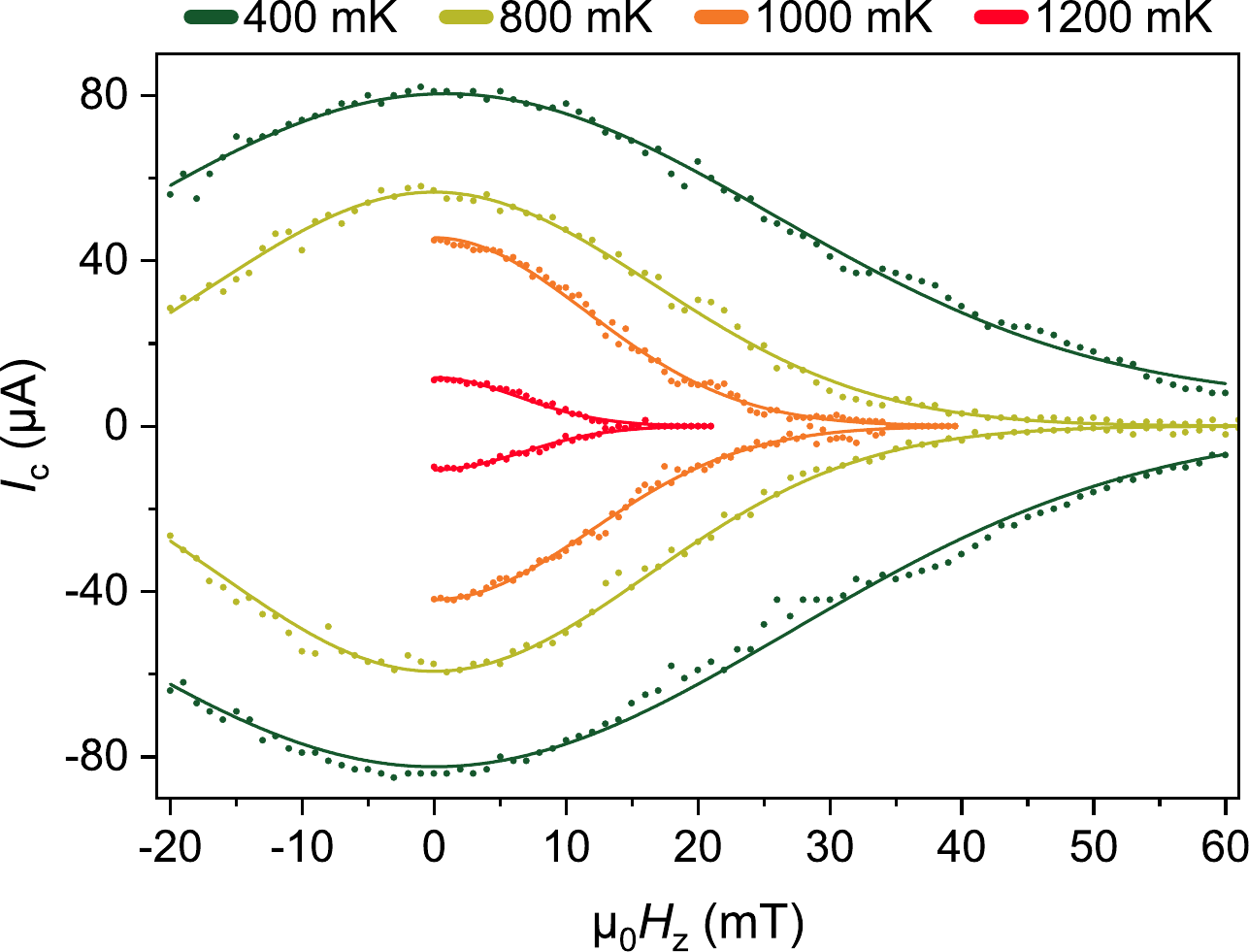}
 \end{array}$}
 \caption{A more detailed $I_\text{c}(B)$ view of the interference patterns obtained on the final stage of ring 2. Here, the points correspond to the measured data, and the line is a Gaussian fit to the data.}\label{Gaussian_decay_SRO}
 \end{figure}

Figure \ref{Gaussian_decay_SRO} shows the $I_\text{c}(B)$-patterns recorded at multiple temperatures on this sample as scatter plot, fitted by a Gaussian decay (solid lines). The data is well fitted by a Gaussian decay. Such $I_\text{c}(B)$-patterns are generally observed on long diffusive junctions~\cite{Cuevas2007,Chiodi2012,Blom2021}. In this limit, the scatter centers lead to the Andreev bound states tracing many different paths through the junction. By the central limit theorem, the phase difference associated with these trajectories will approach a Gaussian distribution, which manifests itself as a Gaussian interference pattern. Since the mean free path of the charge carriers in the normal state is larger than the ring diameter, we do not expect the diffusive limit in our devices. On the other side, we conclude from the time-dependent Ginzburg-Landau simulations presented in Yasui et al.\cite{Yasui2020} that the domain walls are in the extended regime, implying no direct \textit{line of sight} between the superconducting contacts. This might imply that a similar mechanism based on scattering on the sample boundaries is responsible for the Gaussian interference pattern.

\begin{figure}[t!]
 \centerline{$
 \begin{array}{c}
 \includegraphics[width=0.8\linewidth]{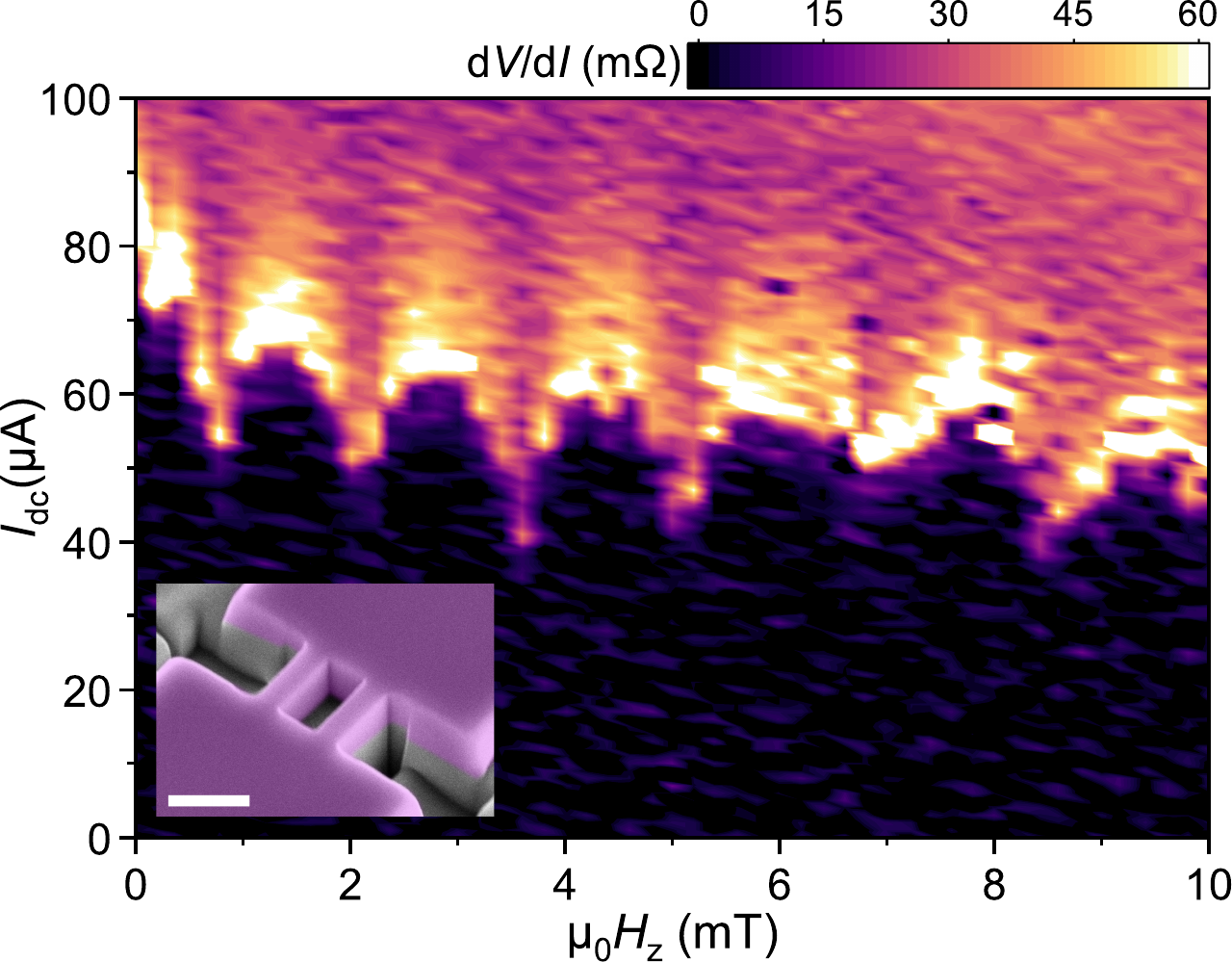}
 \end{array}$}
 \caption{The $I_\text{c}(B)$-pattern, obtained on Ring 4 at 1 K. The inset shows a false colored scanning electron micrograph; the scale bar corresponds to 1~{\textmu}m.}\label{ring4_SQI}
 \end{figure}

 Next, we want to stress that chiral domain walls stabilize in doubly-connected samples, independently of the curvature of the ring. In Figure \ref{ring4_SQI}, we present an interference pattern obtained on Ring 4, which consists of two parallel straight bars between two larger Sr$_2$RuO$_4$ banks (see the bottom row of Figure \ref{C15_table}). We find similar SQUID-like oscillations on this sample as on the other ring samples. This directly rules out any influence of the contact geometry on the stability of the chiral domain walls, and it shows that the chiral domain walls are stable irrespective of the curvature of the arms of the ring.

 \clearpage

\begin{figure}[h!]
 \centerline{$
 \begin{array}{c}
 \includegraphics[width=0.85\linewidth]{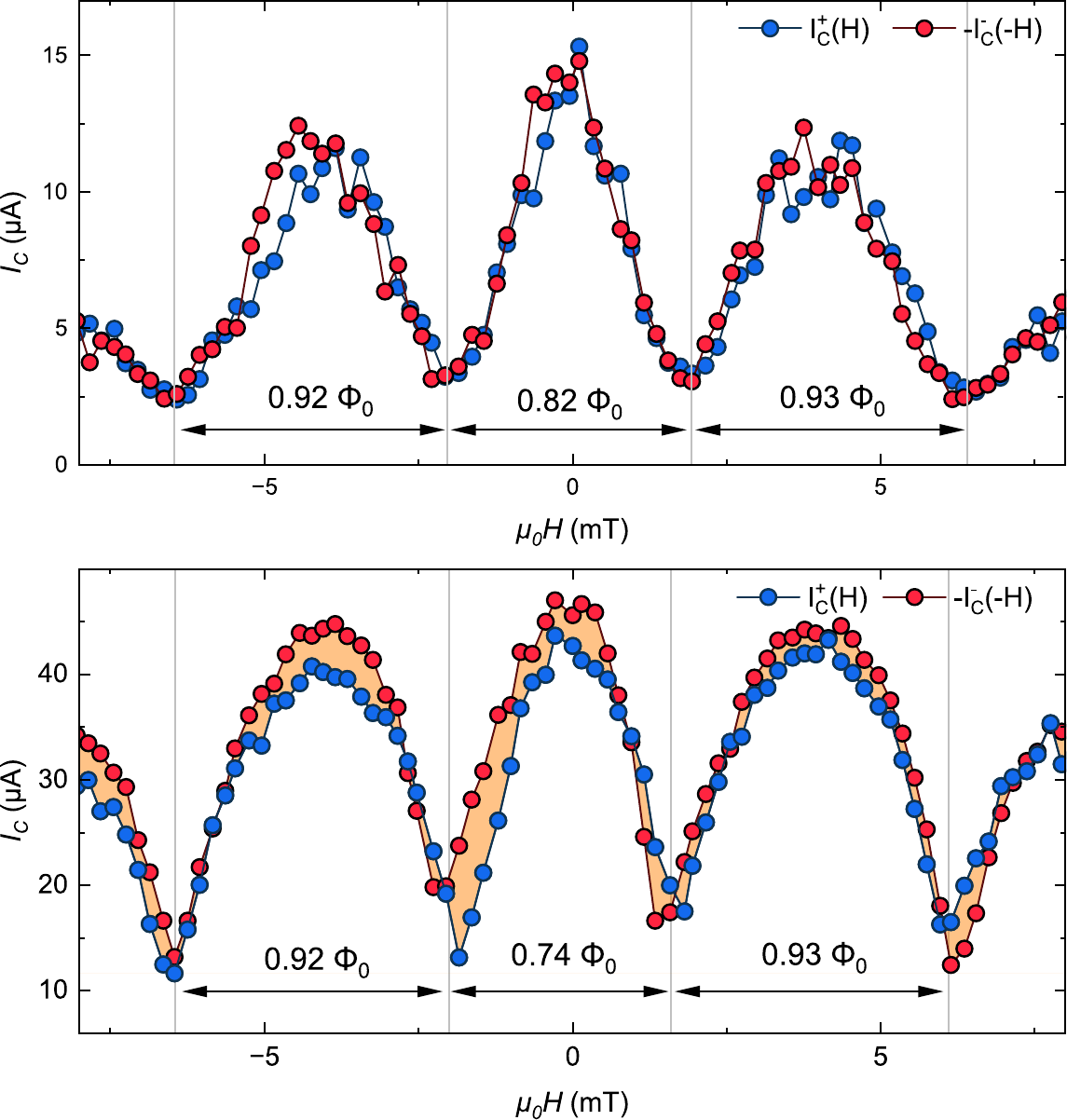}
 \end{array}$}
 \caption{Further TRSB data on Ring 3. (a) Absence of statistically significant TRSB very close to $T_\text{c}$ (at 1200 mK). (b) Emergence of statistically significant TRSB at 1100 mK.}\label{TRSB_B3}
 \end{figure}

 As evidenced by Figure \ref{TRSB_B3}, in addition to the two TRS-breaking samples discussed in the main text, we have found further evidence of TRSB in Ring 3. This ring shows normal SQUID behavior at 1200 mK, but shows three signs of unconventional behavior at 1100 mK: diodicity at zero magnetic field, a narrow central lobe, and a skewed central lobe. By overlaying $I_{\text{c}}^+(H)$ and $I_{\text{c}}^-(-H)$ at 1100 mK, we see that time-reversal symmetry is broken.

\begin{figure}[h!]
 \centerline{$
 \begin{array}{c}
 \includegraphics[width=0.75\linewidth]{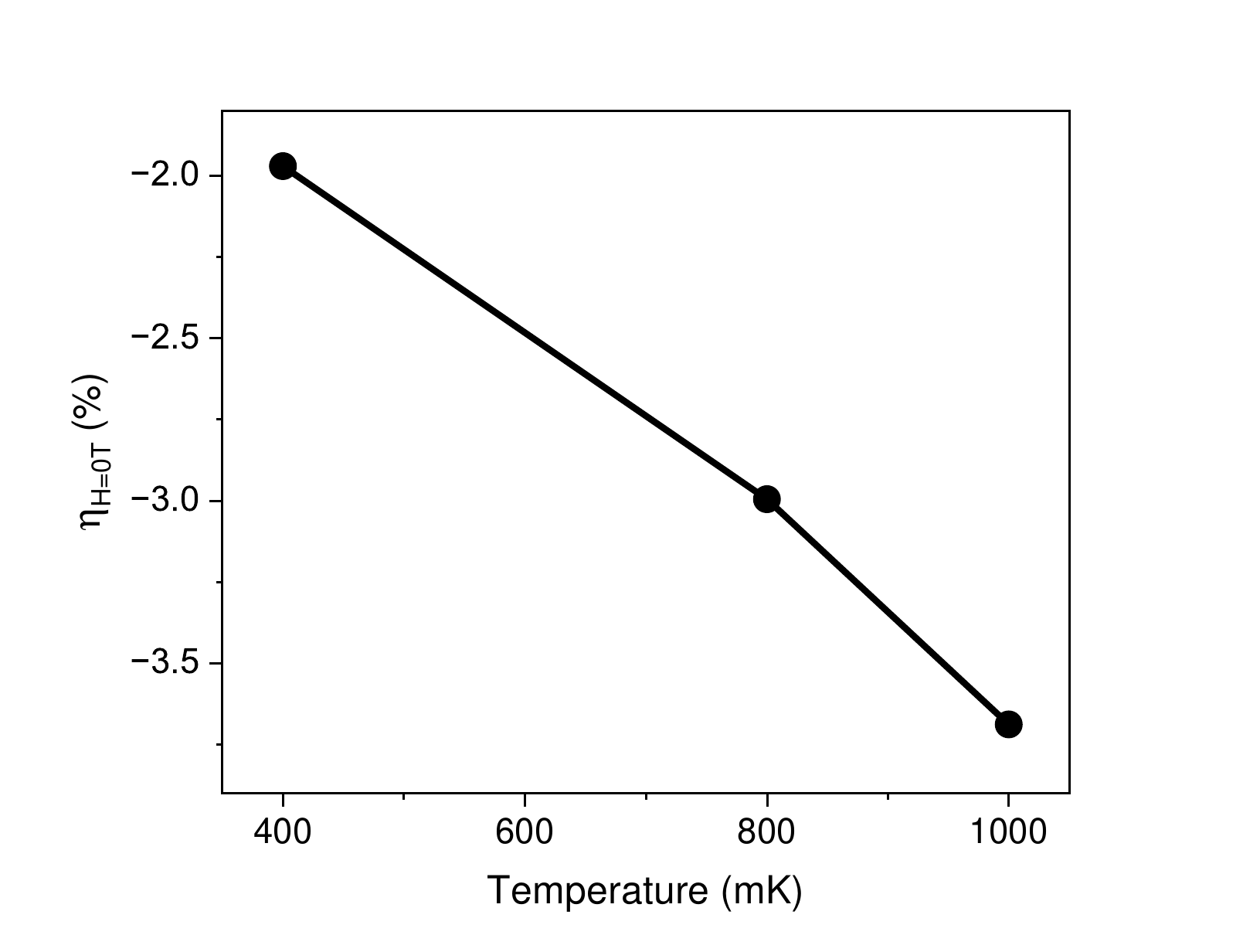}
 \end{array}$}
 \caption{Zero-field diode effect as function of temperature of Ring 2. The diode effect is defined in equation \ref{poep}.}\label{zerofielddiode}
 \end{figure}

 Note that the zero-field superconducting diode effect provides an additional probe for time-reversal symmetry breaking. Therefore, we provide the temperature dependence of the zero-field diode effect of ring 2 in Figure \ref{zerofielddiode}, where we define the diode effect $\eta$ as:

\begin{align}\label{poep}
\eta = \frac{I_C^+ (H) - |I_C^-(H)|}{I_C^+ (H) + |I_C^-(H)|}
\end{align}

\begin{figure}[h!]
 \centerline{$
 \begin{array}{c}
 \includegraphics[width=0.75\linewidth]{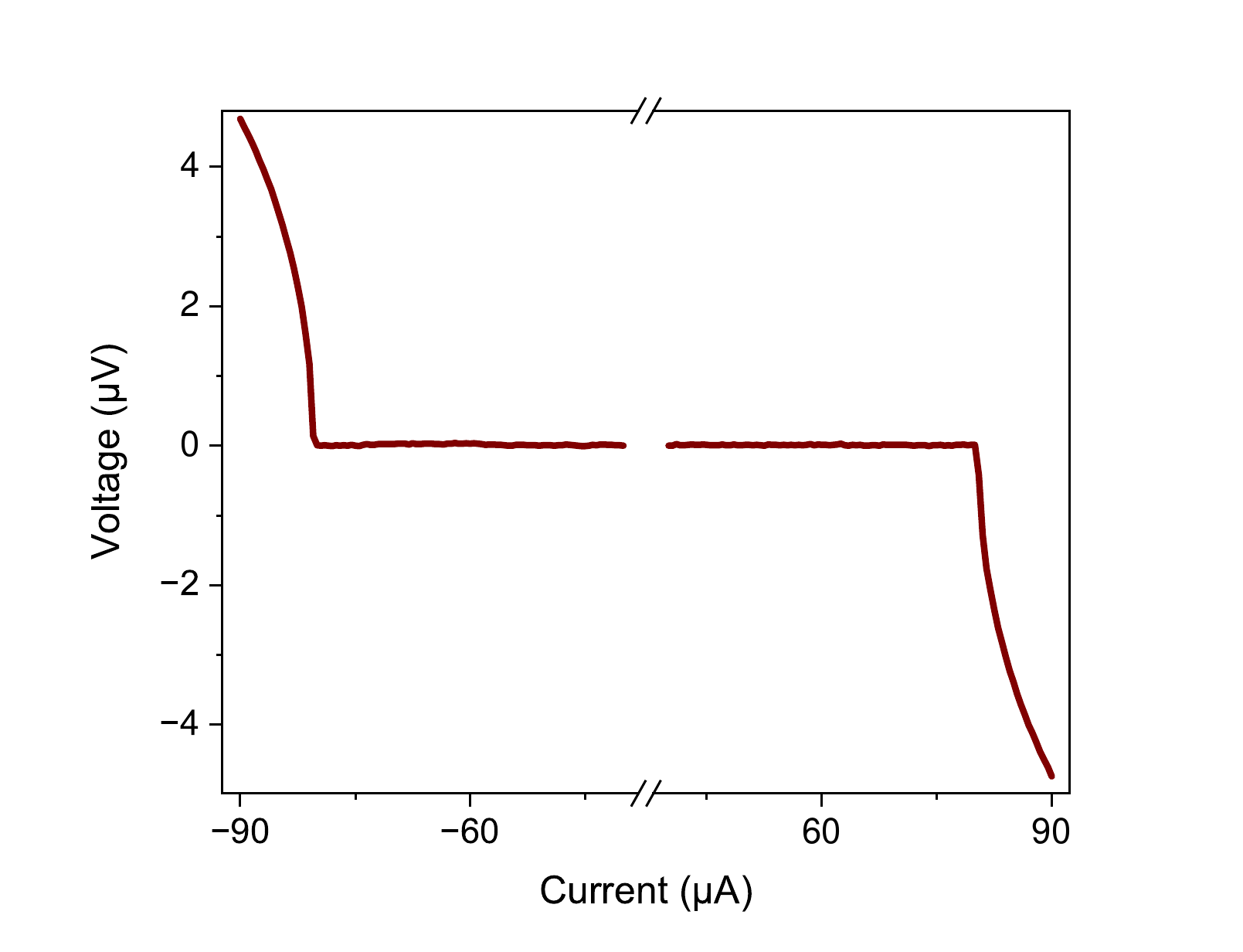}
 \end{array}$}
 \caption{Full-range IV characteristic of Ring 1.}\label{FULLIV}
 \end{figure}

We also append a full-range IV characteristic of ring 1 to complement the data in Figure 1 of the main text in Figure \ref{FULLIV}.

 \clearpage

\clearpage

\section{Extraction of critical current from raw data}\label{How_to_extract}

Key insights into the physics of our devices can be extracted from their critical current oscillations. In this section, we detail the analysis procedure used to extract the critical current from the raw data, minimizing systematic errors and quantifying statistical errors.

\begin{figure}[h!]
 \centerline{$
 \begin{array}{c}
 \includegraphics[width=0.5\linewidth]{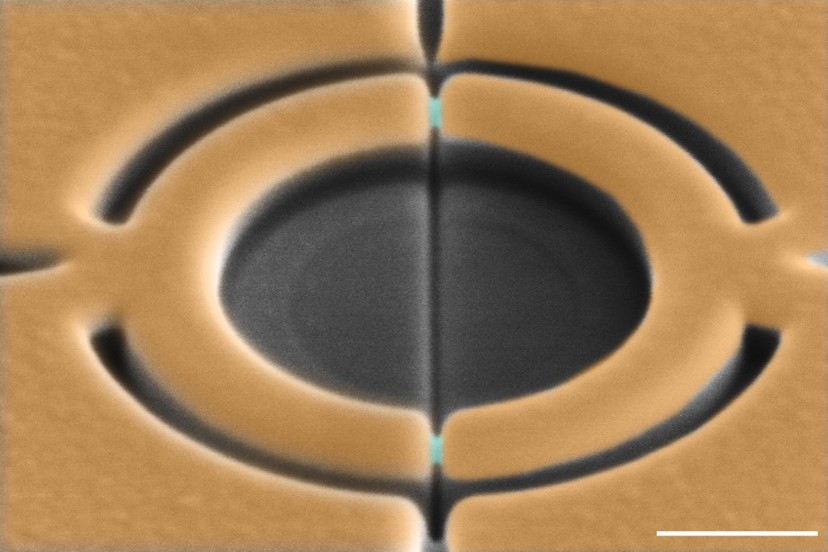}
 \end{array}$}
 \caption{False colored SEM image of the test-SQUID consisting of two SNS junctions. This device is structured with FIB milling out of a MoGe-Ag bilayer (orange). The superconductor is locally removed using line cuts, as highlighted in cyan. The scalebar corresponds to 400 nm.}\label{SEM_MoGe}
 \end{figure}

To illustrate the method, we apply our analysis to a conventional SQUID with two SNS junctions. We fabricated this device from a superconductor/normal metal bilayer and used focused ion beam milling to define the SQUID loop and the junctions by locally removing the superconducting top-layer using line cuts. A false colored SEM image of this test-SQUID is shown in Figure \ref{SEM_MoGe}.

\begin{figure}[h!]
 \centerline{$
 \begin{array}{c}
 \includegraphics[width=0.9\linewidth]{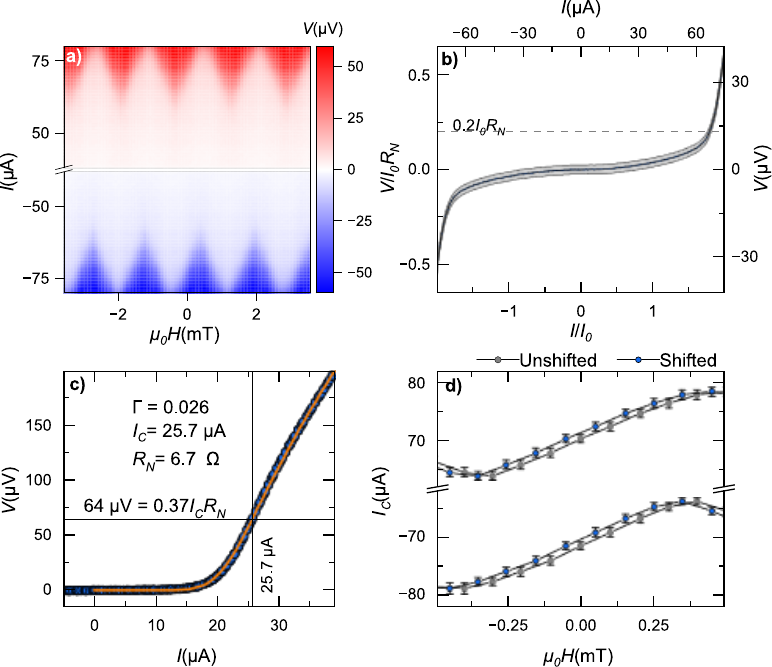}
 \end{array}$}
 \caption{Illustrations of our analysis method using a conventional SNS SQUID as a control experiment. a) The raw dataset used as input for the analysis plotted as a color plot. b) $IV$-data at zero applied magnetic field, after normalization. The statistical error is indicated by the shaded region. c) An IV curve of a single MoGe/Ag Josephson junction fitted by the RCSJ model. Due to rounding effects, a voltage develops far before the critical current. d) Critical current as function of applied magnetic field before and after remnant field subtraction.}\label{data_MoGe}
 \end{figure}

To measure critical current oscillations, we record the $IV$-characteristics at different magnetic fields $H$. This yields the dataset shown in Figure \ref{data_MoGe}a as a color plot. Before extracting $I_C$, we correct for any offset voltage by subtracting the average voltage in a small range of currents around the origin (a few~{\textmu}A is sufficient to cancel out any statistical error). Furthermore, we estimate the stochastic noise in the voltage measurement by analyzing data acquired deep in the superconducting state, where the voltage should be exactly zero, and calculate the standard deviation of the voltage noise.

Next, the data is normalized. Normalization using key device parameters is essential for enabling comparisons to theory, as well as to compare different types of devices. As is conventional in the analysis of SQUIDs, we normalize the currents to the critical current of a single junction $I_0$, and the voltages to the characteristic voltage $I_0 R_N$, where $R_N$ is the normal-state resistance of a single junction. The parameter $R_N$ is extracted from the differential resistance for currents $I \gg I_c$ in a representative portion of the dataset. The critical current is chosen so that the largest critical current for positive current polarity is exactly $2 I_0$. The conversion of voltage to normalized voltage increases the statistical error through standard error propagation techniques. A representative $IV$-curve is shown in Figure \ref{data_MoGe}b.

The normalized $IV$-curves are now used to determine the critical current. The critical current is conventionally determined by a voltage criterion or differential voltage threshold. Typically, the threshold is chosen to be just above the noise level. While this works for superconducting devices with sharp transitions, Josephson-based devices can exhibit rounded $IV$-characteristics due to thermal noise, external (high-frequency) electrical noise, or quantum fluctuations. This rounding, conventionally parametrized using the noise figure $\Gamma$, causes the device to develop a voltage before the critical current. To illustrate this effect, we show an RCSJ-model fit to the $IV$-curve of a single SNS Josephson junction (fabricated using the same techniques as described above) in Figure \ref{data_MoGe}c. To enable better comparison with theory, we use a threshold value of 0.1 $I_0 R_N$ or 0.2 $I_0 R_N$, depending on the degree of rounding in the IV curves.

To extract the critical current corresponding to this threshold, the $IV$-curves are linearly interpolated. Additionally, to propagate the statistical errors from the voltage measurement to $I_C$, we use a parametric bootstrap method. For nearly all analyzed datasets, the resulting uncertainty in $I_C$ is very low. We emphasize that this method only accounts for the uncertainty in measured voltage.

While we have considered the uncertainty in the critical current, we have not yet addressed the error in the applied magnetic field $H$. The dominant source of error in these experiments is the remanent field resulting from the superconducting magnets of the cryostat. Such remanent fields result in a shift of the $I_C (H)$-pattern away from zero magnetic field, inducing an extrinsic asymmetry in the critical current oscillations. This asymmetry could be interpreted incorrectly as evidence of time-reversal symmetry breaking. To compensate for remnant fields, we re-symmetrize the data around the origin by applying a transformation $H \rightarrow H+H_0$ to both the negative and positive critical current branches. Our procedure for the determination of $H_0$ is described to larger extend in section S7.

 \begin{figure}[t!]
 \centerline{$
 \begin{array}{c}
 \includegraphics[width=0.9\linewidth]{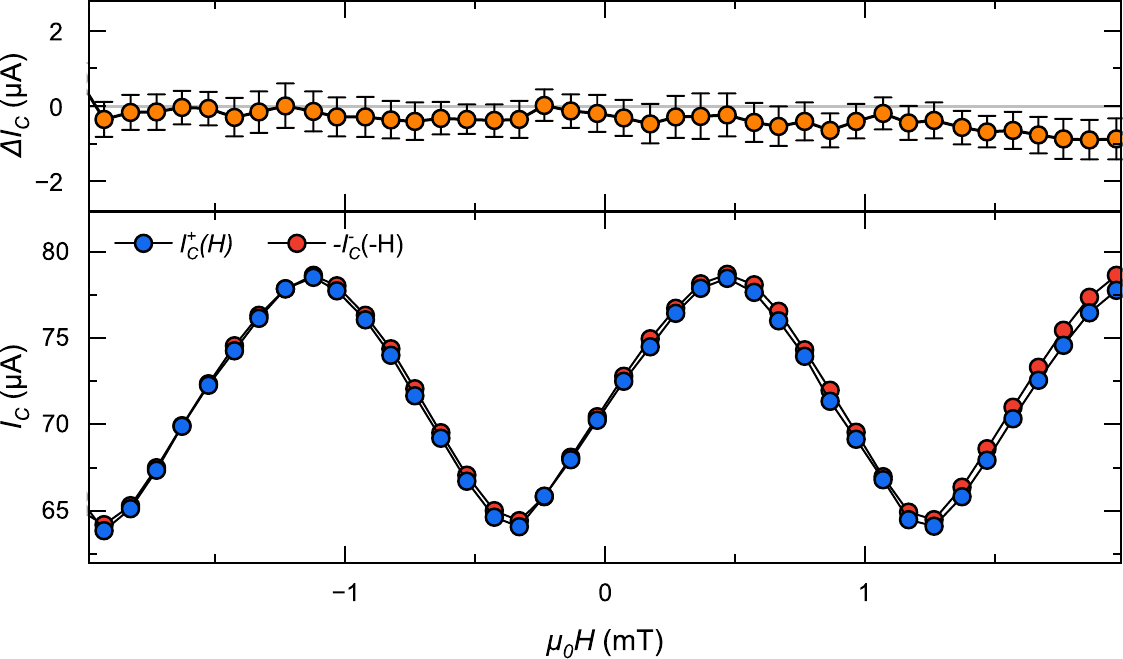}
 \end{array}$}
 \caption{Comparison of $I_C^+ (H$) and $I_C^- (-H)$ to test for time-reversal symmetry breaking. The overlap is within the statistical error for the first three peaks in the $I_C (H)$-pattern, showing the absence of no TRSB in this sample. Note that the apparent diodicity at higher fields is caused by thermal drift effects.}\label{data_MoGe2}
 \end{figure}



After completing all analysis steps, we can visually represent our check for TRSB by assessing the relation $I_C^+ (H)=-I_C^- (-H)$ in Figure \ref{data_MoGe2}. The central four periods show a perfect overlap between the two polarities; for these data points, the difference between the curves $\Delta I_{\text{TRSB}}$ remains well within the statistical margin of error. A larger discrepancy is visible the final two half-periods. We have traced this to a small temperature drift during the experiment.

\clearpage

\section{Toy model for moving chiral domain walls}

We now present a model to heuristically explain the observed reduced central lobe widths in the studied devices. Central to this model is use the observation that external magnetic fields can promote one particular chirality in a chiral superconductors. This effect has been observed in Sr$_2$RuO$_4$ , where the spontaneous magnetization of a sample can be trained by an externally applied field, as observed in Polar Kerr measurements \cite{Xia2006}. Furthermore, in a previous work, we have shown using time-dependent Ginzburg Landau (TDGL) simulations that the position of these chiral domain walls couples to external magnetic fields \cite{Yasui2020}. 

We consider a ring composed of a chiral superconductor, which naturally traps two chiral domain walls in its arms. The domain walls are mobile, but confined to the ring, and make an angle $\phi_D$ with respect to the current direction, as depicted in Figure \ref{toy_model}a. Changing this angle $\phi_D$ keeps most parts of the system’s free energy constant, as a result of the symmetry of the ring. As a first approximation, assuming a fixed length of the CDW as function of $\phi_D$, we can separate the free energy density into two parts:

 \begin{figure}[t!]
 \centerline{$
 \begin{array}{c}
 \includegraphics[width=0.65\linewidth]{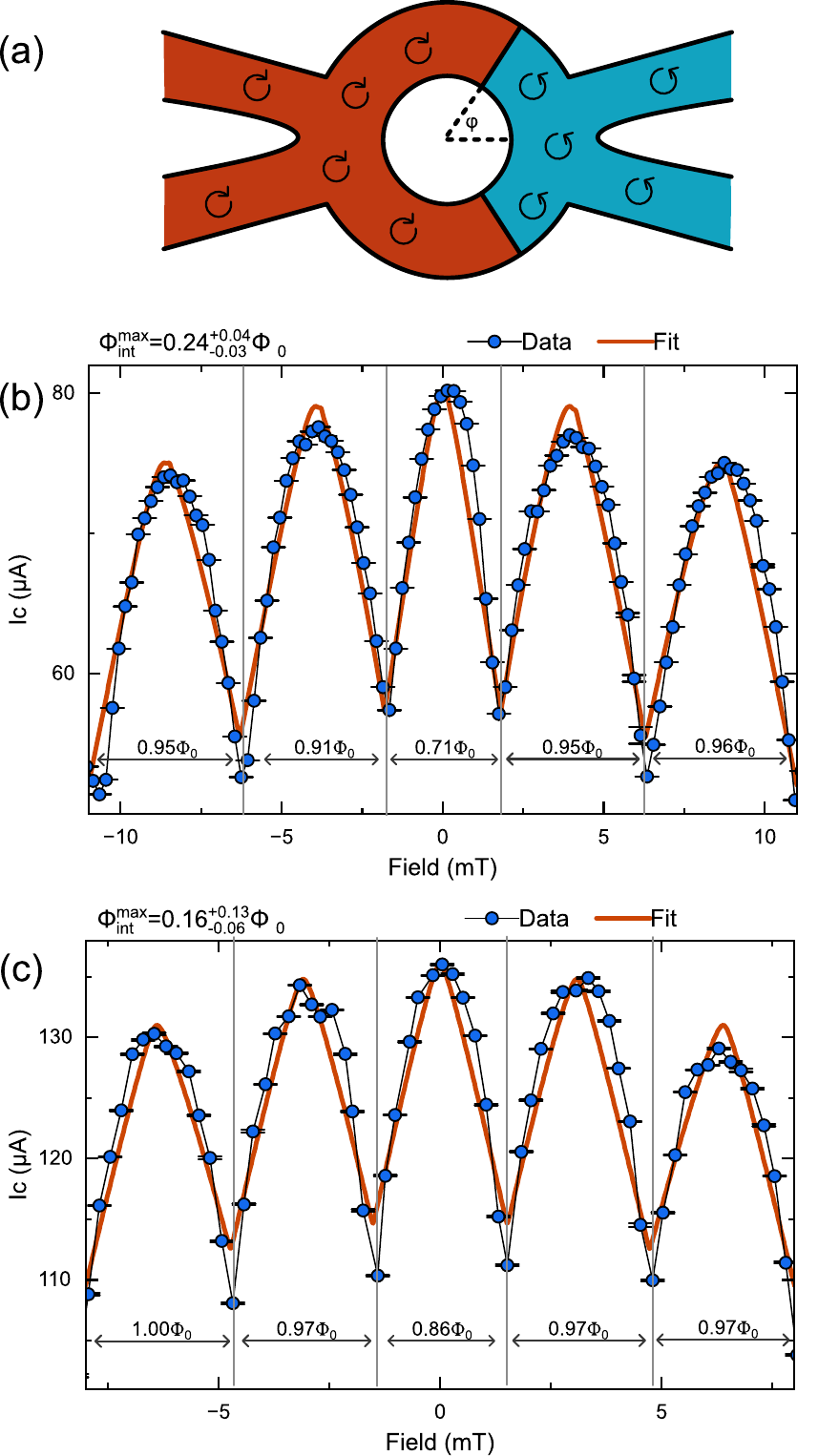}
 \end{array}$}
 \caption{Fitting the $I_C$ oscillations of Ring 1 and Ring 2 to our toy model. a) An illustration of a mobile chiral domain wall in a ring geometry. b) Application of our model on Ring 1 at 800 mK. c) Application of our model on Ring 2 at 800 mK. d Application of our model on Ring 2 at 400 mK. }\label{toy_model}
 \end{figure}
 
\begin{align}\label{eq1}
F = F_0 + F_1(\phi_D)
\end{align}

\noindent where only the second term of eq. \ref{eq1} depends on $\phi_D$. Assuming a linear coupling $\alpha$ between the magnetic field and the chiral order parameters $\Psi_+$ and $\Psi_-$, we obtain:

\begin{align}\label{eq2}
F_1(\phi_D) \propto \int \text{d}V \left[ \alpha(\Psi_+B-\Psi_-B)+B^2 \right]
\end{align}

\noindent where in the last step, we have absorbed some geometric information into $\alpha$. Now expanding $B=H+M(\phi_D)$ and minimizing the free energy reveals a linear proportionality between $\phi_D$ and $H$. 

Naturally, this linear proportionality can only hold for small external fields. As H grows large, the domain wall will reach the neck of the device. Here, it will get pinned, as the two-domain wall configuration is more favorable in energy than a homogeneous ring \cite{Yasui2020}. This prevents the formulation of larger shielding currents. To transition between the low-field mobile domain wall regime, and the high-field static domain wall regime we use a sigmoid function:

\begin{align}\label{eq3}
\phi_D (H) =  \frac{\pi}{2} + \phi_{\text{m}} \frac{1-\exp(\frac{H}{a})}{1+\exp(\frac{H}{a})}
\end{align}

The parameters $\phi_m$ and $a$ parametrize the minimum domain wall angle and the coupling of the domain to the external field, respectively. Although this function does not directly derive from theory, it enables exploration of the effect of a moving domain wall on critical current oscillations. To incorporate the moving domain wall model into our simulations, we add the field generated by a chiral domain to the externally applied flux:

\begin{align}\label{eq4}
\Phi = \Phi_{\text{ext}}  + \Phi_{\text{int}}^{\text{max}} \frac{\phi_D+\frac{\pi}{2}}{\frac{\pi}{2}}
\end{align}

Here, $\Phi_{\text{int}}^{\text{max}}$ is the flux generated by a homogeneous ring of single chirality. This parameter is a fitting parameter as well.

With this extension to the model, we can fit it to the experimental data presented in the main text. In Figures \ref{toy_model} we fit the model to the $I_C$ oscillations of Ring 1 and Ring 2.

\clearpage

\section{Modeling $I_C$ oscillations with the RCSJ model}

\begin{figure}[h!]
 \centerline{$
 \begin{array}{c}
 \includegraphics[width=0.95\linewidth]{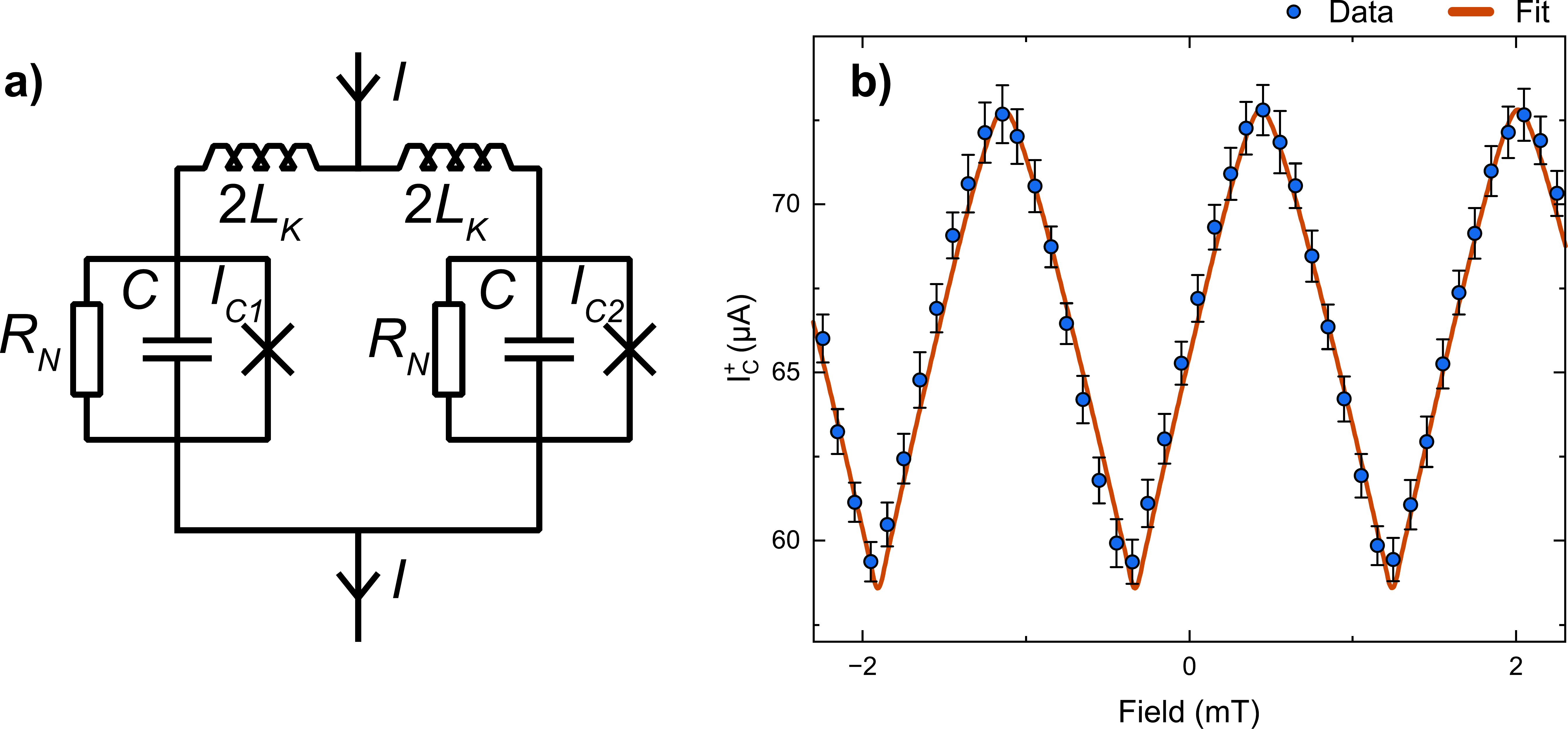}
 \end{array}$}
 \caption{The basis of our fitting models. a) RCSJ model of a SQUID. Our model incorporates asymmetric critical currents, but assumes symmetric $R_N$ and inductance. b) Application of our model on the control sample, showing perfect agreement between RCSJ theory and the experiment.}\label{RCSJ_model}
 \end{figure}

To analyze the proprieties of our devices, we compare the observed critical current oscillations with a theoretical SQUID model. We have interpreted the results in the context of RCSJ model, which is a lumped-element model that treats Josephson junctions as ideal elements with current-phase-relation $f(\delta)$, shunted by resistive ($R_N$) and a capacitive ($C$) channels. A schematic of such a model SQUID is shown in Figure \ref{RCSJ_model}a. In extended junctions such as SNS junctions, the large junction length results in a negligible capacitance. This justifies working in the overdamped ($C \rightarrow 0$) limit. In normalized units, the SQUID of Figure \ref{RCSJ_model}a is described by the coupled differential equations:

\begin{align}\label{eq1_1}
\frac{I}{2} + J = (1-\alpha_I) f(\delta_1) + \dot{\delta} + \beta_c\ddot\delta_1 \\ 
\frac{I}{2} - J = (1-\alpha_I) f(\delta_2) + \dot{\delta} + \beta_c\ddot\delta_2
\end{align}

Here $I$ is the applied current and $J$ represents the circulating current ($J = \frac{1}{2}(I_1-I_2)$). The junctions have normalized critical currents of $1-\alpha_I$ and $1+\alpha_I$ respectively; $\alpha_I$ parametrizes the asymmetry between the weak links. Moreover, $\beta_c$ is the Stewart-McCumber parameter, which characterizes the damping inside the junctions.

 As our junctions have a negligible capacitance, we fix this parameter at 0.001. These equations are coupled through the relation $\delta_2-\delta_1=2π(\Phi_a+\frac{1}{2} \beta_L J)$, where $\Phi_a$ is the externally applied flux and $\beta_L$ is the screening parameter $2LI_0/\Phi_0$, with $L$ denoting the sum of kinetic and magnetic inductance. We do not account for asymmetries in resistance or capacitance. 

By solving these coupled differential equations over a range of bias currents, we can generate IV characteristics for a set of SQUID parameters. We do this by recording the evolution of the phases $\delta_1$ and $\delta_2$ over time, using the \textit{DifferentialEquations.jl} module of the programming language Julia. We can then use the second Josephson relation to calculate the steady-state voltage corresponding to the applied bias current. Restricting this analysis to sinusoidal current-phase-relations $f(\delta)=\sin⁡(\delta)$, we can use this approach to simulate critical-current oscillations for any value of $\alpha_I$ and $\beta_L$.

To compare these simulations to experimental data, we calculate the expected shape of the critical current oscillations for a range of $\alpha_I$ and $\beta_L$ values at sufficiently small spacing. Next, we apply cubic interpolation to generate accurate model oscillations for any value of $\alpha_I$ and $\beta_L$, which are used to fit experimental data. However, before we can compare these oscillations to the data, we incorporate a Gaussian envelope $\exp⁡(\frac{-H^2}{2σ^2})$, which parametrizes the decay of the individual Josephson junctions. The model can then be compared with theory using a Markov chain Monte Carlo fitting procedure. For this we use a Gaussian likelihood function. 

Figure \ref{RCSJ_model}b demonstrates the effectiveness of this procedure in fitting the critical current oscillations of the control SQUID fabricated from a SN-bilayer introduced in section \ref{How_to_extract}, which exhibits a critical current asymmetry. Although the RCSJ model fits the control SQUID very well, the RCSJ model cannot be applied to our Sr$_2$RuO$_4$ rings, due to the narrower central lobe and the TRSB in these samples. 

\clearpage

\section{Josephson energy of a chiral domain wall}

\begin{figure}[h!]
 \centerline{$
 \begin{array}{c}
 \includegraphics[width=1\linewidth]{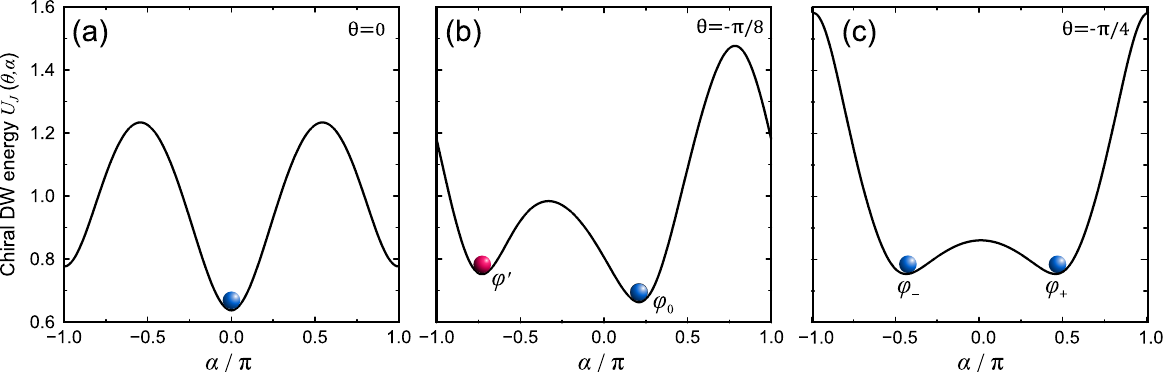}
 \end{array}$}
 \caption{Josephson energy of a chiral domain wall for three different orientations relative to the current flow direction ($\theta$). $\alpha$ is the phase difference between the two sides of the domain wall junction. (a) $\theta = 0$, the junction has a stable phase at $\alpha=0$ and a metastable phase at $\alpha=\pi$. (b) Here $\theta = -\pi/8$; compared to $\theta=0$, the stable state (blue) has a higher energy, while the energy of the metastable state (red) is lowered. The junction has developed arbitrary phases $\varphi_0$ (stable) and $\varphi’$ (metastable). (c) As $\theta \rightarrow -\pi/4$ the two minima continue to approach in energy, and form a degenerate ground state at $\theta = -\pi/4$, with a Josephson phase $\pm \varphi$. The $U_j(\theta, \alpha)$ plots are adopted from Ref. \cite{sigrist_role_1999}}\label{Sigrist_figure}
 \end{figure}

Sigrist and Agterberg \cite{sigrist_role_1999} showed that the Josephson energy of a chiral domain wall is determined by the relative alignment of the two domains with respect to the transport direction (i.e., current flow across the junction). This concept is illustrated in Figure \ref{Sigrist_figure}, where $\theta$ represents the angle between the transport direction and the normal to the plane of the domain wall junction. When the bias current is perpendicular to the domain wall ($\theta = 0$), $U_J(\alpha, 0)$ has one stable ground state at $\alpha=0$, and one metastable state at $\alpha = \pi$. In this case, no spontaneous supercurrent flows across the domain wall, and time-reversal symmetry is conserved. However, as illustrated in \ref{Sigrist_figure}b and c, the two states can change their energy and corresponding phase, depending on the orientation of the domain wall

Figure \ref{Sigrist_figure}b shows the energy of a system where the transport direction has a $\theta = -\pi/8$ angle normal to the ChDW plane. This raises the energy of the stable state (previously at $\alpha = 0$), and lowers the energy of the metastable state. However, more striking is that the ground state has developed an arbitrary phase of $\varphi_0$, which is neither zero nor $\pi$. This phase offset of the ground state results in a spontaneous supercurrent flowing across the ChDW junction in the absence of any external bias, resulting in time-reversal symmetry breaking. In this sense, a domain wall acts as a so-called $\varphi_0$ junction. The energy profile of which has a single ground state with a phase offset of $\varphi_0$. The main difference here is the presence of an additional metastable state $\varphi’$.

As $\theta \rightarrow -\pi/4$, the two states of $U_J(\alpha, \theta)$ continue to approach in energy and ultimately form a degenerate ground state at $\theta = -\pi/4$ (see Figure \ref{Sigrist_figure}c). This energy profile corresponds to that of the so-called $\varphi$-junction, characterized by a double-well ground state with phases $\alpha = \pm \varphi$. Such systems have been realized in Josephson junctions with parallel $0$ and $\pi$ segments, tuned to yield a spatially averaged phase that is neither 0 nor $\pi$ \cite{Goldobin2007, phi_junction_experiment}. A unique characteristic of the $\varphi$-junction is that the $IV$-characteristic can exhibit two distinct critical currents, corresponding to $+\varphi$ and $-\varphi$ states.

\clearpage

\section{Further details on the bistable $I_\text{c}$ states}

\begin{figure}[h!]
 \centerline{$
 \begin{array}{c}
 \includegraphics[width=0.9\linewidth]{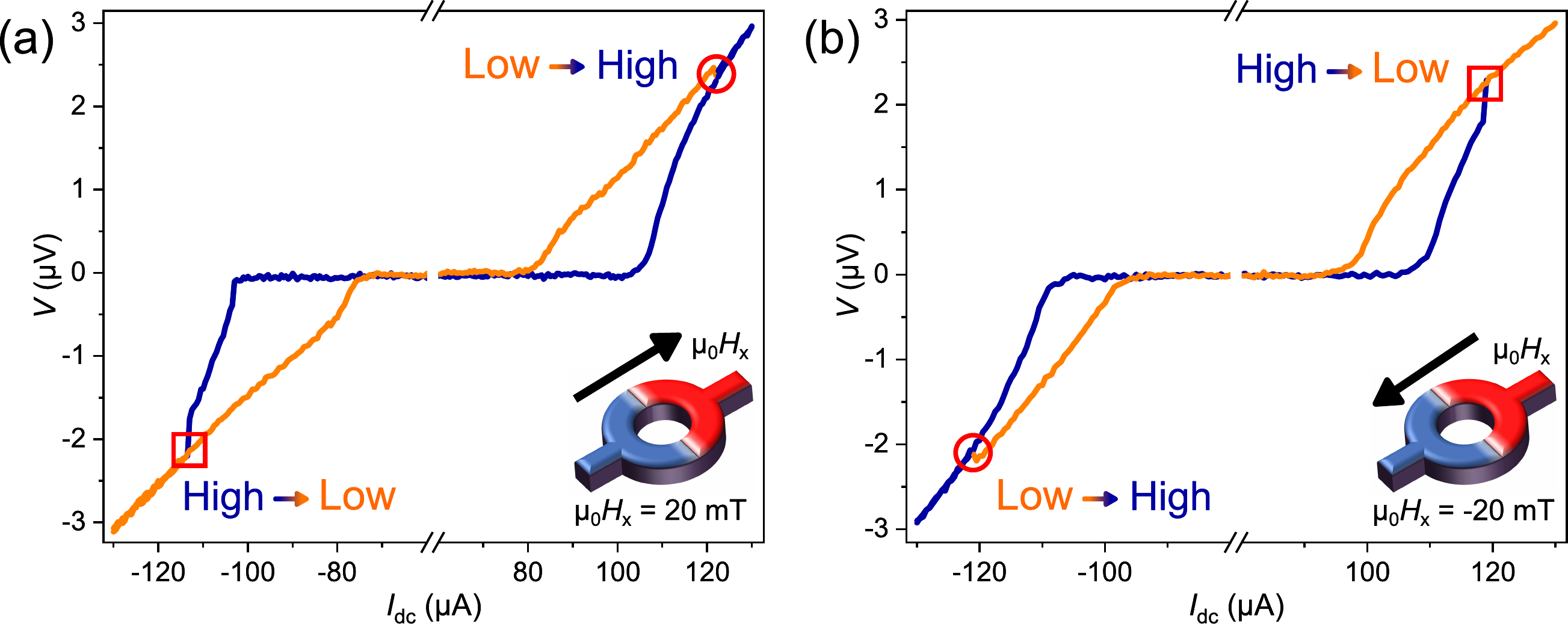}
 \end{array}$}
 \caption{$IV$-characteristics obtained at 600 mK, under the application of an in-plane field of -20 mT along the $x$-direction (line connecting the measurement leads attached to the ring). We observe a similar two-level critical current state as presented in Figure 4 of the main text. Again, we plot the high $I_\text{c}$ state in blue, the low $I_\text{c}$ state in orange, and the $IV$-characteristic for zero field in pink. Note however that roles of $I_\text{+}$ and $I_\text{+}$ in changing between the $I_\text{c}$ states are exchanged when the field direction is reversed. Both $I_\text{+}$ and $I_\text{-}$ are indicated by red circles and the sweeping direction is indicated by black arrows. The inset shows a schematic representation of the respective roles of $I_\text{+}$ and $I_\text{+}$ in changing between the $I_\text{c}$ states, which is opposite as shown in Figure 4 of the main text.}\label{switching_other_field}
 \end{figure}

The bistablility of the critical current states is a universal property of our Sr$_2$RuO$_4$ devices. In Figure \ref{switching_other_field}a, we show that the critical current is split into two branches under the application of an in-plane magnetic field of $\mu_0H_{\text{x}}$ = 20 mT, where switching operations are carried out by the application of a current above a threshold value. This entails that sufficient positive bias current switches the low $I_\text{c}$ state to the high $I_\text{c}$ state, whereas negative bias currents result in the reverse operation. In Figure \ref{switching_other_field}b, we show the results of the reversed field direction, $\mu_0H_{\text{x}}$ = -20 mT. We find that the role of bias current is exchanged when we reverse the field direction.

\begin{figure}[h!]
 \centerline{$
 \begin{array}{c}
 \includegraphics[width=0.6\linewidth]{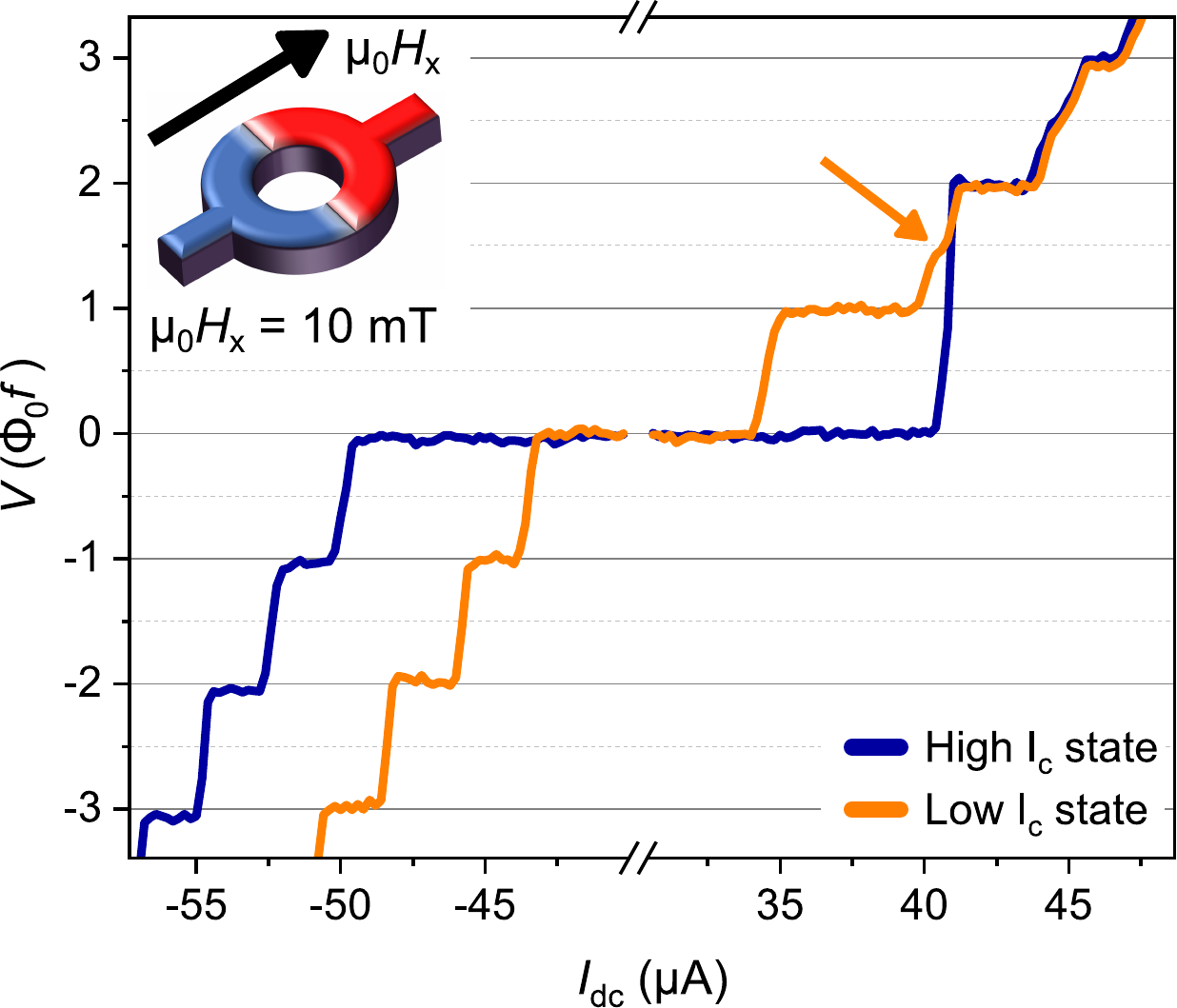}
 \end{array}$}
 \caption{$IV$-characteristic acquired under simultaneous radiation of microwaves (145 MHz) and the application of an in-plane field of 10 mT along the $x$-direction on Ring 1 at 50 mK. After resetting the field (i.e., setting the field to zero and subsequently reapplying the same field), we only find integer Shapiro steps on the negative current polarity. Besides, the first step is lacking on the positive bias polarity and only in the low $I_\text{c}$ state do we observe a half-integer Shapiro step, indicated by the yellow arrow.}\label{shapiro_IP1}
 \end{figure}

As reported in the main text, the application of in-plane magnetic fields can alter the domain structure as well as the response to microwave radiation. Here, we show additional data acquired under the application of microwaves: in Figure \ref{shapiro_IP1} we show the $IV$-characteristics of the low and high critical current state, analogous to the data presented in Figure 4 of the main text, however, after removing and re-applying the magnetic field. We find that reapplication of the magnetic field results in a similar but different $IV$-characteristic, where the half-integer steps are absent on the negative current polarity. On the positive current polarity, we observe a missing first half-step on the low critical current state and even a missing integer step on the high critical current state.

\begin{figure}[h!]
 \centerline{$
 \begin{array}{c}
 \includegraphics[width=0.9\linewidth]{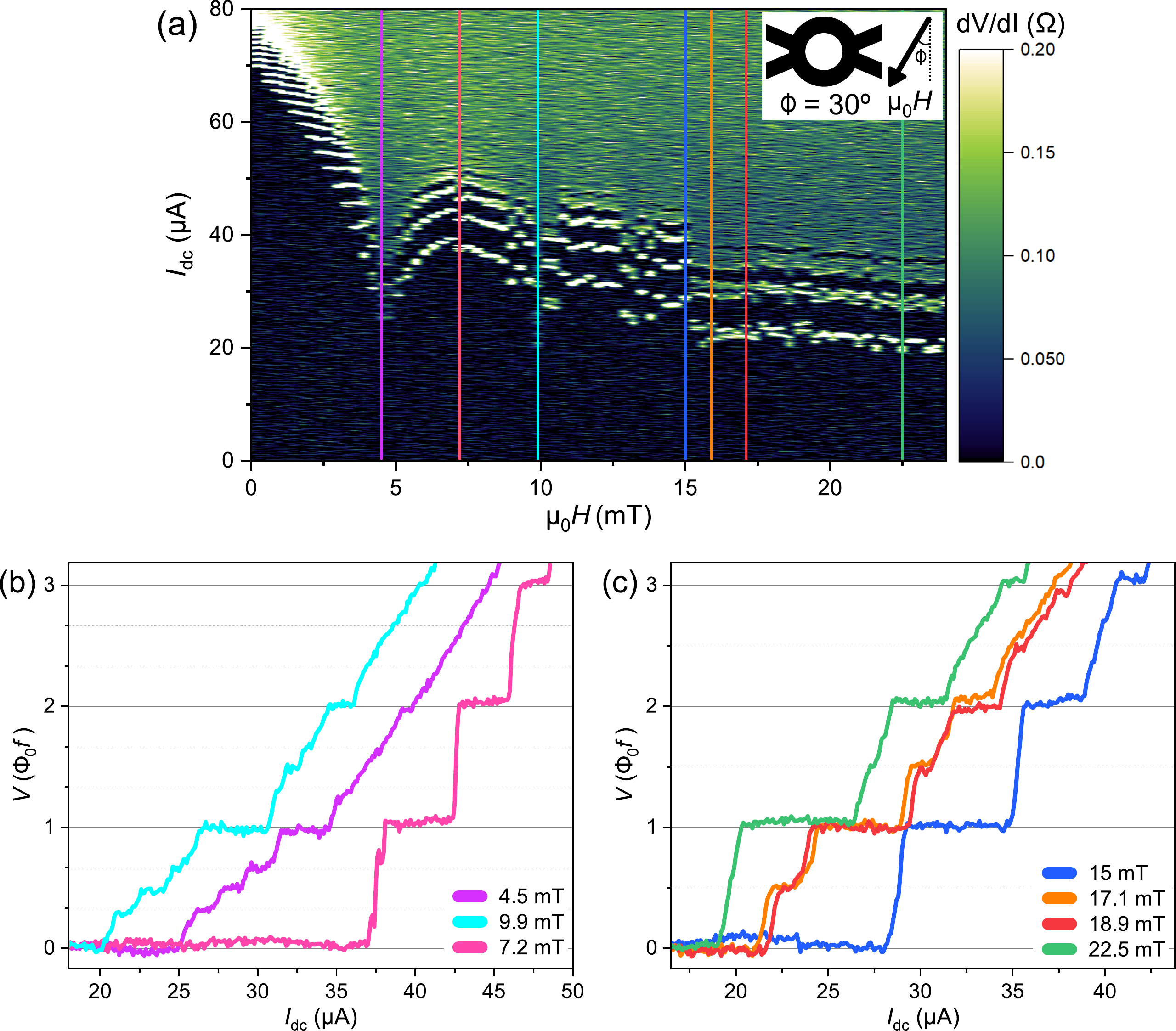}
 \end{array}$}
 \caption{a) In-plane magnetic field sweep obtained under the application of microwaves (145 MHz) at 50 mK. There is a 30-degree angle between the $y$-axis and the magnetic field direction (see inset). To erase any field history, we reduce the field to zero between each field step and we cycle both current polarities twice at each field. In (a) we plot the up-sweep of the second cycle (i.e., increasing current direction). (b) and (c) show $IV$-characteristics extracted from (a). The color coding of the vertical reference lines corresponds to the color of the curves in (b) and (c). Data acquired on Ring 1}\label{shapiro_IP2}
 \end{figure}

In order to track the magnetic field dependence of the fractional Shapiro steps, we perform a magnetic field sweep under microwave radiation (see Figure \ref{shapiro_IP2}). Before this experiment, we verified a reproducible half-integer Shapiro response at finite in-plane fields under an angle of 30 degrees with respect to the normal to the current direction (see inset in Figure \ref{shapiro_IP2}). Interestingly, we observe a Fraunhofer-like interference pattern with Shapiro steps close to $I_\text{c}$. In Figure \ref{shapiro_IP2}b,c we plot $IV$-characteristics at specific fields, extracted from Figure \ref{shapiro_IP2}a. Close to the minima of the interference pattern we find $\frac{1}{3}$-fractional Shapiro steps (i.e., 4.5 mT and 9.9 mT). Naturally, at these minima, the first harmonic of the current-phase relation is suppressed, and therefore it can be dominated by the higher harmonics. This effect is commonly observed in, for example, $0-\pi$ transitions in $SF$-hybrid junctions\cite{Stoutimore2018,Yao2021}. Beyond 15 mT, however, there seem to be no oscillations in the $I_\text{c}(B)$-pattern. In this regime, we see the appearance of the half-integer steps, which might be distinguished from the trivial fractional steps observed in the minima of $I_\text{c}(B)$. At even higher magnetic fields, the half-integer steps disappear, and only integer steps are observed. Besides, integer steps dominate at the maxima in the $I_\text{c}(B)$-pattern.

Although the mobile domain walls and their associated bistability of the critical current are observed universally in all samples, the direction and magnitude of the in-plane field required to trigger the bistable $I_\text{c}$-branches vary between samples. Switching between the critical current states is always accompanied by an apparent hysteresis loop in the $IV$-characteristic. This hysteresis can be used to quantify the difference between the critical current states.

To do so, we define a parameter that measures the absolute voltage difference between the low $I_\text{c}$ state and the high $I_\text{c}$ state. We calculate this parameter by summing the absolute voltage difference between the two states at each current:

\begin{align} \label{H7eq1}
\text{Hyst.} = \sum_{I} |V(I_{\rightarrow}) - V(I_{\leftarrow})|
\end{align}

\begin{figure}[h!]
 \centerline{$
 \begin{array}{c}
 \includegraphics[width=1\linewidth]{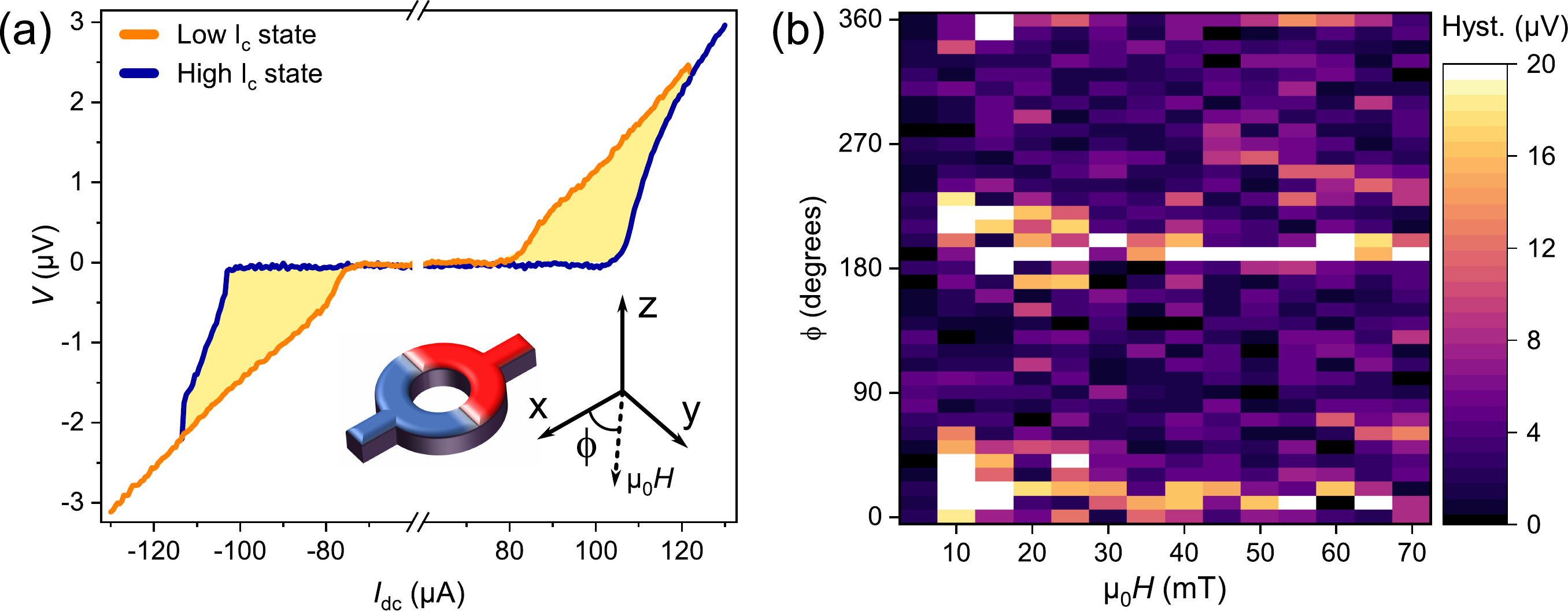}
 \end{array}$}
 \caption{Hysteresis parameter (summed difference between the high $I_\text{c}$ state and the low $I_\text{c}$ state) extracted for different combinations of field magnitude and direction. Note that each pixel represents a single $IV$-characteristic (sweeping back and forth). Here $\phi$ is the polar angle in the xy-plane (with the x direction along the ring, as defined in Figure \ref{switching_other_field}). Note that no hysteresis is observed at in-plane magnetic fields below 10 mT. At $\phi= 20$ degrees we find clear hysteresis behavior for a large range of magnetic fields, indicating switching between different $I_\text{c}$ states. Before setting each field and starting the $IV$-measurement, the field was reduced to zero.}\label{stability_SRO1}
 \end{figure}

Here we sum over each current probed in the $IV$-characteristic, $V(I_{\rightarrow})$ is the voltage measured during the forward sweep, and $V(I_{\leftarrow})$ is the voltage measured on the reverse sweep. Naturally, this parameter gives a measure for the surface enclosed by the hysteresis loops (see Figure \ref{stability_SRO1}a): it is zero if there is no hysteresis and the two critical current states are equal. However, it is finite if any hysteresis exists between the current and retrapping current, which indicates a switch between the critical current states.

In Figure \ref{stability_SRO1}b we plot the hysteresis parameter for a large combination of field magnitudes and directions for Ring 1. Here $\phi$ is the polar angle in the $xy$-plane, and $\phi = 0$ corresponds to the direction along the ring, as defined in the inset of Figure \ref{stability_SRO1}a. We find $\phi= 20$ degrees the optimal angle to produce the switching behavior for this sample. On the other hand, Ring 2 showed the most pronounced bistable $I_\text{c}$ property at $\phi= 55$ degrees.

\clearpage

\section{Determination of the Remanent Field Through Data Analysis}
As described earlier, one critical source of uncertainty that dominates our analysis is the possibility of a remanent field. It is well-established that superconducting magnets trap vortices, especially when used at higher fields, which introduce a remanent field in the experimental setup even when all field-generating electrical currents are turned to zero. In our data, such a remanent field yields a horizontal shift in the interference patterns which could be easily misconstrued as evidence of TRSB.

Offsetting a remanent field is notoriously difficult, as active cancellation requires a very carefully calibrated zero-field sensor (e.g. a well-calibrated Hall probe). In this work, we take a different approach where the remanent field is found in the analysis by examining the symmetry of the $I_\text{c}(H)$-patterns.

 \begin{figure}[h!]
 \centerline{$
 \begin{array}{c}
 \includegraphics[width=1\linewidth]{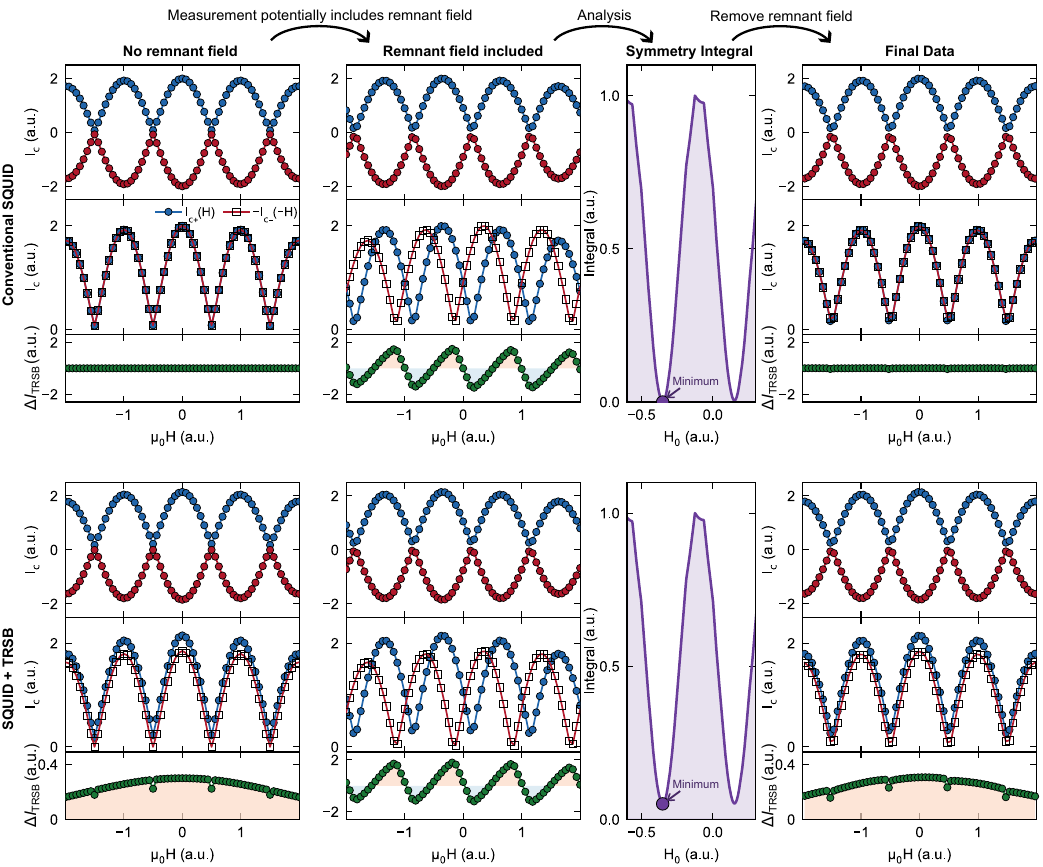}
 \end{array}$}
 \caption{Illustration of our remanent field cancellation method, on computer-generated example dataset. (a-d) Show the procedure performed on a conventional SQUID that preserves TRSB. A remanent field in the measurement procedure results in a finite $\Delta I_{\text{TRSB}}$, which is erroneously interpreted as time-reversal symmetry breaking. Through minimization of the integral in equation \ref{eq1_1}, the remanent field can be removed from the dataset and $I_{\text{TRSB}}$ equals the situation without remnant field. (e-h) Show the procedure performed on a SQUID exhibiting a zero-field diodicity. In this case, the remanent field enhances the apparent TRSB. By minimizing the asymmetry integral, we regain the true time-reversal symmetry breaking parameter $I_{\text{TRSB}}$.}\label{remanent_illustration}
 \end{figure}

As has been strongly emphasized in the main text, the key indicator of TRSB is the identity $I_\text{c+}(H) \neq I_\text{c-}(-H)$. A remnant field erroneously leads to this inequality. In that case, the integral

\begin{align}\label{eq1_1}
\int \left[ I_\text{c+}(H) - |I_\text{c-}(-H)| \right]^2
\end{align}

\noindent must also be non-zero. The larger the asymmetry, the larger the value of the integral will be. Conversely, the integral will be zero if the remanent field is zero. This integral is therefore a direct measure of the time-reversal breaking combined with the remnant field.

Mathematically, a remanent field $H_0$ acts on an interference pattern by shifting both the positive and negative critical current branch away from the origin by $H_0$: $H\to H-H_0$. Therefore, we can undo the remanent field shift by finding the value of $H_0$ and subsequently performing the inverse transformation. To find the remanent field shift, we minimize the above integral:

\begin{align}\label{eq1_2}
\arg\min_{H_0} \int \left[ I_\text{c+} (H + H_0) - I_\text{c-}(H + H_0) \right]^2
\end{align}

We illustrate this procedure Figure \ref{remanent_illustration}, which considers two computer-generated example datasets. The first is a conventional SQUID, with time-reversal symmetry preserved, in which case $I_\text{c}(\Phi)$ can be modeled as $2I_0 |\cos(2\pi \Phi/\Phi_0)|$. The second is a SQUID that breaks time-reversal symmetry through a diodicity: 

\begin{align}
I_{\text{c}\pm}(\Phi) = \pm 2I_0 |\cos(2\pi \Phi/\Phi_0)| + I_\alpha \exp(-\Phi^2/2\sigma^2)
\end{align}

\noindent where the strength and shape of the diodicity is encoded in $I_\alpha$ and $\sigma$. The first column of Figure \ref{remanent_illustration} shows the two example computer generated datasets discussed above: $\Delta I_{\text{TRSB}}$ is zero for the conventional SQUID, but non-zero for the TRSB SQUID, indicating time-reversal symmetry breaking only in the latter. In case these devices are measured in a system exhibiting a remnant field, both devices show a non-zero $\Delta I_{\text{TRSB}}$, falsely suggesting time-reversal symmetry breaking in both. By carrying out the procedure described above and minimizing the integral of equation \ref{eq1_1}, we remove the remnant field from the data and any $\Delta I_{\text{TRSB}}$ results purely from time-reversal symmetry breaking.
\clearpage


%